%% file: n6530.tex
\documentclass[referee]{aa} 
\usepackage{graphicx}
\usepackage{longtable}
\usepackage{psfig}
\usepackage{amsmath}
\usepackage{deluxetable} 
\usepackage{aas_macros}
\usepackage{amssymb}
\usepackage{natbib}
\bibpunct{(}{)}{;}{a}{,}{,}
\def\ngc#1{NGC$\,$#1}

\def\vmi{\hbox{\it V--I\/}}
\def\bmv{\hbox{\it B--V\/}}

\def\Min{${}^{\prime}$\llap{.}}
\def\Sec{${}^{\prime\prime}$\llap{.}}
\def\deg{${}^\circ$}
\def\min{${}^{\prime}$}
\def\sec{${}^{\prime\prime}$}
\begin{document}
   \title{The Star Formation
   Region \ngc6530: distance, ages and Initial Mass Function 
   \thanks{Based on observations made with the European Southern Observatory
     telescopes obtained from the ESO/ST-ECF Science Archive Facility.}}

   \author{L. Prisinzano \inst{1,2}, F. Damiani \inst{2}, G. Micela \inst{2}
    and S. Sciortino \inst{2}}

   \offprints{loredana@astropa.unipa.it}

   \institute{
Dipartimento di Scienze Fisiche ed Astronomiche, Universit\`a
di Palermo, Piazza del Parlamento 1, I-90134 Palermo Italy
\and INAF - Osservatorio Astronomico di Palermo, Piazza del Parlamento 1, 90134
Palermo Italy}
\date{Received March 16, 2004; accepted September 28, 2004}
\abstract{We present astrometry and $BVI$ photometry, down to $V\simeq22$,
    of the very young open cluster 
   \ngc6530,  obtained from
observations taken with the Wide Field Imager camera at the MPG/ESO 2.2\,m
Telescope. Both the $V$ vs. $B-V$ and the $V$ vs. $V-I$ color-magnitude
diagrams (CMD) show the upper main sequence dominated by very bright cluster 
stars,
while,  due to the high obscuration  of the giant molecular
cloud surrounding the cluster, the blue envelopes of the diagrams at $V\gtrsim 14$  
are limited to
the main sequence stars at the distance of \ngc6530. This  particular
structure of the \ngc6530 CMD allows us to conclude that its distance 
 is about $d \simeq 1250$\,pc, significantly lower than the previous   
determination of d=1800\,pc.\\
  We have positionally matched our optical catalog with the list of X-ray
sources found in a Chandra-ACIS observation,
 finding a total of 828 common stars,
90\% of which are pre-main sequence stars in \ngc6530. 
 Using evolutionary tracks of \citet{sies00}, mass
and age values  are inferred for these stars. The median age of the cluster 
is about
 2.3\,Myr; in the mass range (0.6--4.0)$\,M_\odot$, 
the Initial Mass Function (IMF) shows a power law
index $x=1.22\pm0.17$, consistent with both the Salpeter index (1.35),
  and with
the index derived for   other young clusters ;
 towards smaller masses the IMF shows a peak and then 
it starts to decrease.
   \keywords{ Open clusters - individual: \ngc6530 - photometry - astrometry -
membership - pre-main sequence stars - Luminosity and Initial Mass Function}}

\titlerunning{\ngc6530} \authorrunning{Prisinzano et al.}
\maketitle 
\section{Introduction\label{intro}} 
Star formation regions and very young open  clusters are crucial systems
for  understanding   the star formation process because they allow us to derive
  Initial Mass Functions that are  not affected by stellar and/or dynamical
evolution effects.   
\ngc6530 ($l=6.14, b=-1.38$) is an example of a 
very young open cluster  located in front of    
 the M8 giant molecular cloud, also referred to as Lagoon Nebula \citep{lada76} . 
 The brightest part of M8
is the Hourglass Nebula, illuminated by the extremely young O star 
Herschel
36 (O7 V). Other stars of spectral type O, 9 Sgr (O4 V) and the binary  system
HD 165052 (O6.5 V + O6.5 V) excite the Lagoon Nebula.
The open cluster \ngc6530 is located about   9\Min8~ eastwards  of the
 Hourglass Nebula (corresponding to a projected distance from the Nebula of
3.6 pc, using the cluster distance $d=1250$\,pc  estimated in this work) 
 and can be recognized by its  brightest stars.

Since the work of \citet{walk57}, 
several investigations have been devoted to study this cluster and to estimate
its parameters. Using mainly photoelectric and photographic  observations,
 many attempts have been done to estimate
the distance of \ngc6530 from the  CMDs, which show a
normal cluster main sequence down to about A0 stars; fainter stars lie 
above the main sequence,   indicating that the cluster is so young  
that its low mass members are still gravitationally contracting
\citep{walk57}. The lack of the complete cluster main sequence made it hard to 
obtain reliable estimates of the cluster distance. 
 \ngc6530 distance has been estimated by several authors to be in the range
(1300--2000)\,pc (cfr. Table \ref{distance_tab}), by taking advantage
of spectroscopic observations, to determine spectral types, and proper motions,
to select cluster members. 
\begin{table}[t]
\centering
\caption {Literature distance values   for \ngc6530}
\vspace{0.5cm}
\begin{tabular}{ccc} 
\hline
\hline
Distance (pc)& $({\rm m-M})_0$	&Reference \\
\hline
1380--2000	&10.7--11.5&\citet{walk57}\\
1300		&10.6	   &\citet{the60} \\
1400		&10.7	   &\citet{walk61}\\
1580		&11.0	   &\citet{hilt65}\\
1780		&11.25	   &\citet{alte72}\\
1380		&10.7	   &\citet{kila77}\\
1820		&11.3	   &\citet{saga78}\\
1900		&11.4	   &\citet{chin81}\\
1860		&11.35	   &\citet{mcca90}\\
1600--2000	&11.0--11.5&\citet{anck97}\\
1400	        &10.74     &\citet{lokt97}\\
1800		&11.25	   &\citet{sung00}\\
560--711	&8.75--9.26&\citet{lokt01}\\
\hline
\hline
\end{tabular}
\label{distance_tab}
\end{table} 

  On the contrary,
$UBV$ observations  have  allowed several authors to study the
reddening law and to derive  cluster reddening values very similar to
the average value $E(B-V)=0.35$  recently derived by \citet{sung00}, who
have assumed   a foreground reddening value, $E(B-V)_{fg}=0.17$, as estimated by 
\citet{mcca90}. 
Finally, the age of the cluster  was  estimated to be 
in the range (1.5--2.0)\,Myr by several
authors \citep{alte72,saga78,sung00}, while \citet{dami04} give a median age of 
0.8\,Myr. Due to the young age of this cluster, low mass stars are
expected to be still in pre-main sequence  phase,
 as already found in \citet{sung00}
by using photometric data down to the limiting magnitude $V\sim 17$.
 119 X-ray point sources in the Lagoon Nebula region
have been recently detected by \citet{rauw02} 
 in  a 20ksec {\it XMM--Newton} observation; they found that
 most of the X-ray sources are associated with pre-main sequence stars of
 low and intermediate mass. However, a larger list of point sources 
in the same region, with a much better spatial resolution, has been
 recently obtained  by \citet{dami04}, using Chandra ACIS-I X-ray data.
Damiani et al.  show that 
their source sample is made for at least $90\%$ by cluster members and that
out of the 884 X-ray sources found, only 220 have a counterpart in the
\citet{sung00} optical catalog and most of   this latter 
 are pre-main sequence 
members of \ngc6530.
As suggested by Damiani et al., the remaining X-ray sources are likely stars
with magnitude fainter than $V\sim17$.
The lack of published photometric data at magnitudes fainter than $V\sim17$ has 
motivated the analysis of deep optical images  to study the 
low mass stellar population in the \ngc6530 field  
and, in particular, the low-mass 
pre-main sequence stars of the cluster by means of
  cross-correlation with available X-ray data.  

In this paper, we first present 
the observations and the 
data reduction procedure (Sect. 2)
and the CMDs obtained using
the optical catalog (Sect. 3).
 Therefore,
 we present  the cross-correlation of optical and X-ray data, 
 from which membership is determined (Sect. 4), 
 and 	     the cross-correlation of optical and 2MASS IR data (Sect. 5);
mass and age determination of cluster members and the analysis of the
spatial distribution are discussed in Sect. 6. 
 Finally, the Luminosity and  the Initial Mass Functions  (Sect. 7)
 and our  conclusions (Sect. 8) are presented.
%
\section{The Observational Data  \label{datared6530}}
The data used in this work come from the combination of optical {\em BVI}
 images
taken with the Wide Field Imager (WFI) camera at the 
2.2\,m Telescope of the European Southern Observatory (ESO),
 a  60 ks Chandra ACIS
X-ray observation \citep{dami04}
 and public near infrared data from  the All-Sky Catalog of Point Sources
 of the Two Micron 
All Sky Survey (2MASS)\footnote{a joint project of 
the University of Massachusetts and the Infrared Processing and
Analysis Center/California Institute of Technology, funded by the National 
Aeronautics and Space Administration and the National Science Foundation}
\citep{cutr03} available on the WEB\footnote{http://irsa.ipac.caltech.edu/}.
\subsection{Optical Observations and Data Reduction}
\label{optical_data}
The optical observations, consisting of  9
{\em BVI} images of 
  \ngc6530, were taken  using the
WFI camera mounted  at the Cassegrain focus of the 
 ESO 2.2\,m Telescope 
at La Silla (Chile). This instrument consists of a
$4\times 2$ mosaic of CCDs of 2048$\times$4096 square pixels 
with a scale of  
0.238 arcsec/pixel;
each chip is 8\Min12 $\times$ 16\Min25, while the
full field of view (FOV) is   
$34\times 33$ square arcmin.
The observations, retrieved from the ESO/ST-ECF Science Archive Facility,
are part of the  ESO Imaging Survey (EIS) in the context of the
PRE-FLAMES program \citep{moma01}. Details of the observations are given in
Table \ref{obs6530}. The frames were taken under slightly variable
seeing 
conditions with an improvement during the night
of the Full Width Half Maximum (FWHM) of the
point-spread function (PSF)
 from 1.76 to 0.83 arcsec,
as measured on the frames. 
\begin{table*}
\centering
\tabcolsep 0.1truecm
\caption {Log-book of the optical  observations.}
\vspace{0.5cm}
\begin{tabular}{ cccccccc} 
\hline
\hline
\multicolumn{1}{c}{Target}&
\multicolumn{1}{c}{RA (J2000)}&
\multicolumn{1}{c}{Dec (J2000)}&
\multicolumn{1}{c}{Night}&
\multicolumn{1}{c}{Filter}&  
\multicolumn{1}{c}{Exp. Time} &
\multicolumn{1}{c}{seeing} &
\multicolumn{1}{c}{Airmass}\\  
EIS name & (h m s)  & (d m s) & & &[s] & FWHM[\sec] &\\
\hline\\
\ngc 6530 & 18 04 48.0 &$-$24 19 41.0  & 27--28 Jul 2000 & {\it B}&
$1\times30+2\times240$  
& 1.36--1.76 &1.268--1.236 \\
(OC31) &" &" & " & {\it V} & $1\times30+2\times240$   & 1.19--1.36 &1.216--1.194\\
&" &"&"& {\it I} & $1\times30+2\times240$   & 0.83--1.05 & 1.175--1.154\\
\hline
\hline
\end{tabular}
\label{obs6530}
\end{table*}

\begin{figure*}[!ht]
\includegraphics[width=15cm]{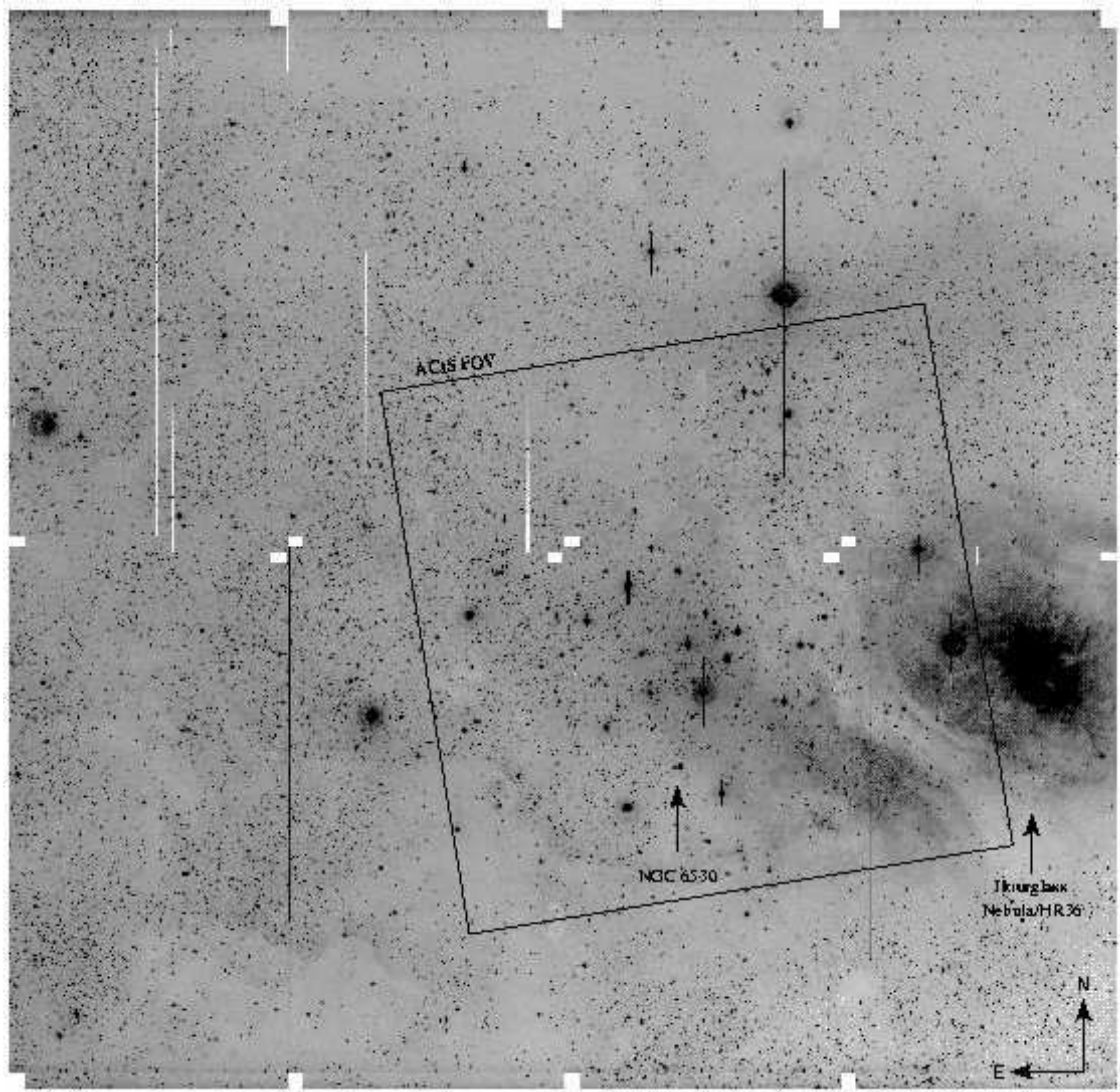}
\caption{Image obtained from the combination of the two deep dithered
  $I$ band images of the region around \ngc 6530 observed with the WFI.
    White squares are unexposed regions due to the deterctor gaps.}
\label{0133_0134}
\end{figure*}

Fig. \ref{0133_0134} shows  the image obtained 
from the combination of the two deep dithered  $I$ band images of
the region around \ngc 6530, corrected for the instrumental signatures as 
described below. The  FWHM of the PSF in this frame is about 0.83 arcsec.
The very young cluster \ngc6530 can be recognized by  
its brightest stars located about 9\Min8 eastwards of the O star 
Herschel 36, embedded in the Hourglass Nebula.  
 The image shows regions of high and low stellar 
surface density according to the irregular pattern
of the molecular cloud   absorption.

 The first stage of the data reduction process has been the 
instrumental calibration of the images through   
the {\tt mscred} package, a
mosaic specific task implemented as an IRAF\footnote{Image Reduction and
Analysis Facility} package for the {\it NOAO\footnote{National Optical Astronomy
Observatory}
Mosaic Data Handling System}.
First the instrumental electronic bias of the images
was subtracted using the 
overscan region and then   
the  images were  trimmed  to remove the overscan region.
Flat fielding for each filter
was  performed using a set of sky
flat fields scaled to the median value of all chips
combined into a master flat using the {\tt mscred flatcombine} 
task. 

A special treatment was  required  for the images in the $I$ band,
affected  by strong effects of fringing. 
To remove this instrumental artifact,  
the fringing pattern, provided by the MPG/ESO 2.2\,m Telescope 
team,\footnote{available at  
http://www.ls.eso.org/lasilla/Telescopes/2p2T/E2p2M/WFI}  
was subtracted scaling it to each exposure.
 
Detection of sources recorded in the analyzed digital frames was obtained
using the DAOPHOT II/ALLSTAR  
photometric routines described in \citet{stet87}. 
A more accurate  
PSF fitting photometry was performed by submitting all  images
to the routine
 ALLFRAME \citep{stet94} that simultaneously uses the geometric and
photometric information from all  images to derive a self-consistent set of
positions and magnitudes for all detected sources.

In the present case, all images are characterized by
a very strong sky background gradient and thus  
particular attention was 
devoted to estimate the minimum sky level  (DAOPHOT parameter {\tt LOW GOOD
DATUM}=50 in unit of standard deviations)
in order to include legitimate
sky pixels where the background is faint. 

A non standard procedure, specifically suited for this case,
 was applied to the chip containing the  star
Herschel 36 because a significant number of spurious detections was 
found due to the emission from the nearby Hourglass Nebula.
 In order to recognize true star detections,
the list of the stars released by ALLSTAR was filtered according to
the {\tt sharp}
parameter. 
After this selection was  applied, the simultaneous 
PSF fitting photometry was  performed with the routine 
ALLFRAME which releases  a
star-subtracted image from each frame. The star-subtracted images
were used to  obtain a 
median image, that was then smoothed with a Gaussian
smoothing $\sigma$ larger than
 the FWHM of the PSF but smaller than the angular scale of the
structures in the nebula. Finally a median image without nebula was 
obtained by subtracting the smoothed star-subtracted median image from the
median image obtained by the original images.
A more reliable star list, determined using the latter median
nebula-subtracted image, was   given as input to ALLFRAME to obtain
the final PSF fitting photometry.

 In order to
convert the profile-fitting photometry to the standard
photometric system, a magnitude zero point was   calculated 
for each
chip. This was  done using the growth curve method  described  in 
 \citet{stet90}.
 First, we   have 
selected the 5268 stars used to define the PSF model; next, all
other objects were removed from the frames and aperture photometry was
carried out at different radii. The DAOGROW code  \citep{stet90} was   
used to
derive growth curves and  COLLECT was used to 
calculate the "aperture correction" coefficient for each chip, from 
the difference between PSF-fitting magnitudes 
 and aperture photometry magnitudes of the selected stars.

The procedure described above was  also applied to a set of images of the
\citet{land92} standard fields  SA92 and SA107 
obtained with WFI/2.2\,m during the same night.

Using the $v$, $b$ and $i$  instrumental magnitudes and the $V$, $B$ and $I$ 
magnitudes  of the Johnson-Kron-Cousins photometric system
of the standard stars falling in these fields,
the transformation coefficients  to the standard system
were performed using the following  equations:
\begin{eqnarray}
\label{scemo}
 v &= V + A_0 + A_1 \times  X + A_2 \times (\vmi), \nonumber\\
 b &= B + B_0 + B_1 \times  X + B_2 \times (\bmv), \\
 i &= I + C_0 + C_1 \times  X + C_2 \times (\vmi). \nonumber   
\end{eqnarray}  
where  $X$ is the airmass and  
$A_0$, $B_0$ and $C_0$ are the magnitude zero points,  
$A_1$, $B_1$ and $C_1$ are the extinction coefficients, while
$A_2$, $B_2$ and $C_2$ are the color terms.
The list of standard stars used to determine the coefficients
contains the standard stars defined in the \citet{land92}
catalog and a number of secondary standard stars present in the Landolt
fields and defined by \citet{stet00}.
 Of these stars, we have between 25 and 161 measurements for each chip,
except for  chip  \# 53   (the north-western chip in Fig. \ref{0133_0134}) 
because neither Landolt's catalogue standards  
nor     Stetson's catalogue standards were falling in this chip. 
Since the observations of the
standard  stars do not cover a wide airmass range, we were not able
to derive the extinction coefficients. 
The extinction coefficients typical of La Silla 
(Stetson, private communication)
were adopted, and 
the zero point and the color term were calculated for all chips
except for chip  \# 53 .
 Stars with magnitude residual values above the 
$3\,\sigma$ level were not  considered 
 for the coefficient fitting.  
   Considering the homogeneous characteristic of the instrument, we 
    choose 
 to compute new zero points fixing the color term to the mean value
 of the color terms of the chips  \# 50, \#51, \#54, \#55, \#56
 for which the number of measured standard
 stars was larger than 40.
Differences smaller than 0.1 mag among the   zero points of different chips
were found. Based on this latter result, we  used their
average   and the above determined  average  color term to 
calibrate the stars in chip  \# 53, where no standard stars
are found. 
 The resulting coefficients  
 are given in Table \ref{calib6530}.
\begin{table}[!htb]
\centering
\caption {Coefficients of the transformation to the standard system for 
each filter and averaged over all chips.}
\vspace{0.5cm}
\begin{tabular}{ccccccc} 
\hline
\hline
Filter& Av. Zero Point&Extinction &Av. Color Term \\
\hline
$V$    &$0.905\pm   0.021$& 0.14 &$0.066 \pm 0.046$&\\
$B$    &$0.528\pm   0.030$&0.25	 &$-0.245\pm 0.019$&\\
$I$    &$1.904\pm   0.009$&0.09	 &$-0.133\pm 0.034$&\\
\hline
\hline
\end{tabular}
\label{calib6530}
\end{table} 

Using these coefficients and the inverted form of  
equations (\ref{scemo}), the {\em B, V, I} magnitudes of 
 all objects detected in at least two filters  
were calculated. 
In order to  test the self-consistency of the photometry, we 
compared the photometry of the stars common to contiguous chips. 
Common stars between chips were found because
the deep mosaic
images where dithered by about 30 arcsec in RA and about 60 arcsec in Dec 
from one another,
in order to cover the inter-chip gaps.
By matching  contiguous chips we  found  about 10--30 common stars  
per chip pair for which we  compared the photometry.
We found  mean offsets in the photometric zero points $\lesssim 4\%$ in 
$V$ and $I$, and $\lesssim 7\%$ in $B$;  this supports the choice to
use the average zero point of the other chips to calibrate the data
in chip \# 53.

In order to consider only  stellar-like objects, 
a data selection was applied by  
    including only   
 objects with  {\tt sharp} between -0.5 and 0.5 \citep{stet87} i.e.
 objects with a brightness distribution consistent with  a point source.
After this selection, our  catalog includes   53\,581 objects.
\subsubsection{Astrometry}  
Celestial coordinates of the detected optical stars were obtained
using the Guide Star Catalogue Version 2.2.01 (GSC\,2.2,  STScI, 2001)
as reference catalogue. 
The first step was  to match the list of celestial coordinates of  
stars retrieved from the GSC\,2.2 catalog, with the list of pixel coordinates 
obtained for each chip  
as described in the previous section. The transformation between the two systems
was  obtained applying 
the appropriate projection to the celestial coordinates and  
a linear transformation to the pixel coordinates. The initial estimate for the
linear transformation is determined using three stars whose 
  celestial and pixel  coordinates  are both known.
  To find the best astrometric solution, 
we matched the two 
catalogs using a conservative 
matching tolerance of 0\Sec3. With this matching tolerance  
 2671 
  common stars were
 found (IRAF task {\tt ccxymatch}) over a total of 
 33\,554 
  sources  (appearing in at least $50\%$ of our frames) 
   used to match the GSC\,2.2 catalog. 
 The matched lists of pixel and celestial coordinates  
 of each chip  were used
 to compute the plate solution. The sky projection was  obtained  using a
 combination of the tangent plate projection and polynomials 
 (IRAF task {\tt ccmap}).  Finally, using the derived astrometric solution,
  we computed celestial coordinates of our photometric catalog stars.

In order to estimate our astrometric accuracy we matched stars of our catalog
with $V<20$, with the
GSC\,2.2 catalog, used as reference to find the astrometric solution and
we  considered the offset
distribution  within a relatively large value (5\sec). From this distribution we
subtracted the expected distribution of  spurious identifications,
obtaining the distribution of  true identifications only. The resulting
 rms is  0\Sec4, that is therefore our
 final accuracy.
 \subsubsection{Comparison with previous catalogs}
 As an external check of our    photometry    
we  selected stars of our 
 catalog  with $V<17.5$ and  compared them with the  \citet{sung00} catalog
 limited to $V \le 17$. 
By matching the two catalogs, we  found a systematic offset in both  
right ascension and declination of $\sim$-1\Sec05 and $\sim$0\Sec34, respectively.

  To estimate the appropriate matching radius for comparing the two catalogs,
we first removed the coordinate offsets and then   considered the offset
distribution within 7\sec. We found that the
rms width of the distribution is 0\Sec5 and that the
 contribution of spurious
identifications becomes dominant at offsets larger than 1\sec.

  With the aim to include only reliable matches between our and the 
Sung et al. catalogs, we 
adopted a matching radius of 0\Sec5, finding 576 stars with  offsets   
-0\Sec01$\pm$0\Sec30 and  
-0\Sec02$\pm$0\Sec22, 	   
in RA and Dec, respectively. 
  The mean offsets and the  rms of the photometric 
residuals are
$0.006\pm0.060$ mag,
$0.007\pm0.075$ mag  and
$0.037\pm0.055$ mag   
in $V$, $(B-V)$, and $(V-I)$,  respectively. These values are 
computed after applying a 3$\sigma$ clipping to
the photometric residuals. 

 Finally, we compared our photometric catalog with the catalog of 
\citet{walk57} and
\citet{kila77} consisting of a total of 319 stars. Of these stars, only the 150 stars
with $V>9.5$,  falling in the WFI FOV and 
with  celestial coordinates   found and 
retrieved from  the SIMBAD database,  were considered.
With a matching radius of 9\sec\, we found  131 stars plus a 
number of mis-identifications
that we  discarded based on their photometric residuals. 
By considering the stars having  
both   $B$ and $V$ magnitudes in the \citet{walk57} and 
\citet{kila77} catalog,   we found
$\Delta V=0.002 \pm 0.073$  and 
$\Delta (B-V)=0.006\pm 0.081$, after applying
a 5$\sigma$ clipping to the photometric residuals.    
The remaining 19 stars   with
no  counterpart in our catalog have $V\sim 10$,
hence they are saturated  in the WFI frames.

 All previous comparisons  show that our  magnitudes and colors are
in excellent agreement with those measured by other authors. Table \ref{wfi_sung_wk}
shows the results of such comparisons;
col.s 1--9  list the sequential number, 
celestial coordinates and photometry 
of our catalog; the star number used by 
\citet[ denoted as $ID_{SCB}$]{sung00} and the astrometric and photometric residuals
obtained from the comparison with the \citet{sung00} catalog are given in col.s 10--15;
the star number used by 
\citet{walk57} and \citet[ denoted as $ID_{WK}$]{kila77} and the 
astrometric and photometric residuals
obtained from the comparison with the \citet{walk57} and \citet{kila77} catalogs
 are given in col.s 16--20. 
\input{wfi_sung_wk}
%
\subsection{X-ray and near-infrared  data}
To better understand part of our optical data we 
    used a list of X-ray sources,
 presented in \citet{dami04}. These data were obtained
from a 60 ks observation carried out on June 18-19, 2001 with the Chandra 
ACIS-I CCD detector having a very narrow PSF (FWHM$\sim$0.5\sec~ on-axis).
 The observation consists of a 
FOV of 17\min$\times$17\min~ pointed 
toward RA=$18^h04^m24^{\rm s}$\llap{.}38, Dec=-24\deg21\min05\Sec8, i.e.
 centered
on the \ngc6530 cluster.  
 884  point-like X-ray sources were detected as described in detail by
\citet{dami04}. 

In addition, to complement the optical data, 
we  used $JHK_S$ infrared (IR)  photometry taken
from the   All-Sky Point Source public catalogue
of the Two Micron All Sky Survey (2MASS) \citep{cutr03}.
\section{The Color-Magnitude Diagrams}
\subsection{Optical data}
\label{cmd_optical_data} 
The final optical photometric catalog has been obtained by adding to
our data, the stars of the \citet{sung00} catalog that were not detected
in the WFI images. These are  
  very bright stars ($V\lesssim 10$ mag), saturated
in our images, or  stars falling out of the WFI FOV or on 
the edges of the chips, for a total of 123 objects.
To select these stars, we considered the stars
of the \citet{sung00} catalog whose
 astrometric offset with respect to our catalog 
is larger than 1\sec,
the radius for which the number of matched stars drops to   
zero.  

CMDs of all stars of the complete optical catalog
were constructed after selecting  the
  stars with errors in the $V$ and $I$ magnitudes both 
 smaller than  0.2 mag; stars with larger
(statistical) errors  cannot be placed accurately on the CMD
 so their inclusion is meaningless. We  applied this   
selection  only
to the $V$ and $I$ magnitudes because in the following analysis we mainly
will use the $V$ vs. \vmi~ CMD. Nevertheless, the
$V$ vs. \bmv~ diagram has been obtained using stars of the cleaned catalog with
errors in $B$ smaller than  0.2 mag. 
 Using this selection criterion on our data we   have found 
  that our detection limit
is $V \sim 23$.    
Completeness of our data was determined via artificial star tests. 
A total of about 1650 artificial stars
were added on each chip and photometry of the artificial frames was performed
using the same  reduction procedure applied to the original frames. 
In order to test the completeness of the data used in the following analysis,
the results  were obtained by considering the
number of retrieved artificial stars with  {\tt sharp} between -0.5 and 0.5 and
errors in the $V$ and $I$ magnitudes
 smaller than 0.2 mag.
 We   have found 
  that our data are $100\%$ complete above $V\sim20$ and $I\sim18.5$,
 while they are more than $50\%$ complete above $V\sim22$ and $I\sim19.5$.

\begin{figure*}
\includegraphics[width=10cm]{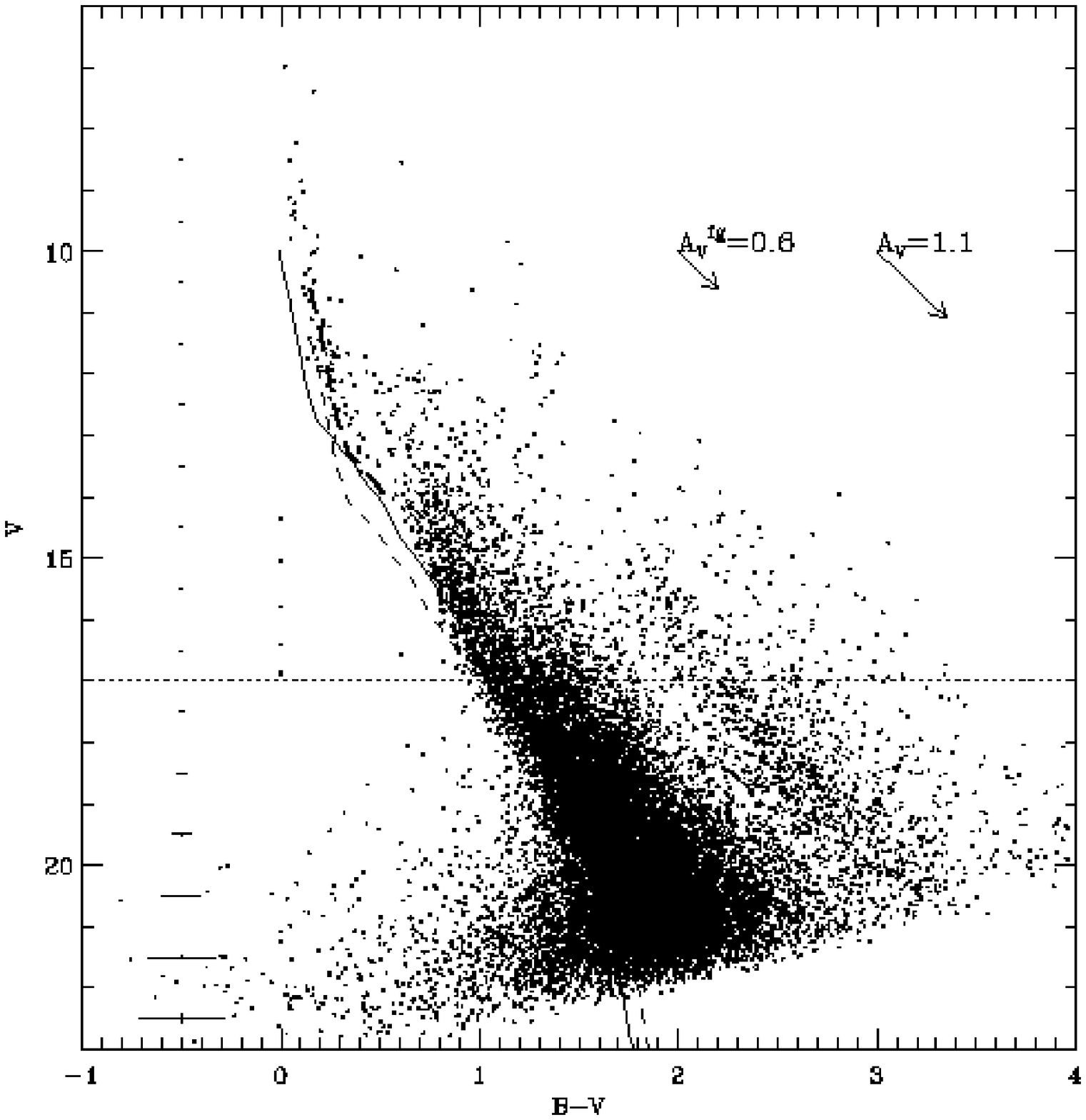}
\includegraphics[width=10cm]{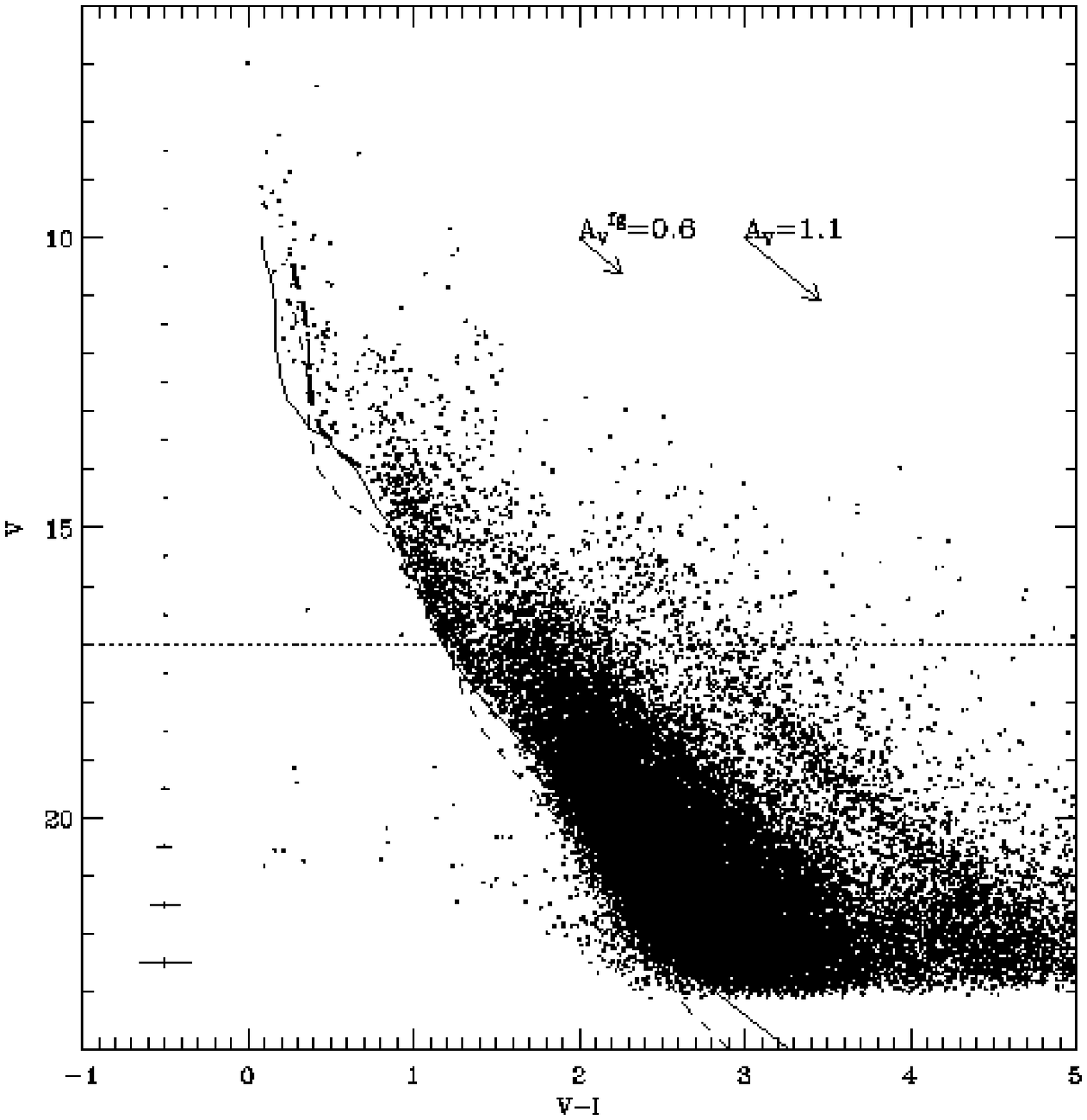}
\caption{The   $V$ vs. $B-V$  and the $V$ vs. $V-I$
CMDs of 
  selected  stars (20\,253 and 37\,553, respectively) 
 in the WFI FOV.
Horizontal bars indicate the median errors in color, while vertical bars
(barely visible)
indicate the median errors in  magnitude for bins of one magnitude.
 The horizontal dashed line in each diagram 
indicates the magnitude   
completeness  limit of the \citet{sung00} catalog. 
The {\it solid line} is the Zero Age Main Sequence (ZAMS)
of \citet{sies00}   assuming the
cluster distance  $d=1250$\,pc and $E(B-V)=0.20$; the 
{\it thick long dashed line}
is the same ZAMS at the same distance but with the 
average cluster reddening  $E(B-V)=0.35$
 estimated by \citet{sung00}; finally the {\it dashed line} is the ZAMS 
 of \citet{sung99} at the 
  distance of 1800 pc and  the 
 cluster reddening $E(B-V)=0.35$, derived by Sung et al. (2000).
 The extinction 
vectors $A_{V}^{\small \rm fg}=0.6$ and $A_V=1.1$
correspond to  reddening $E(B-V)=0.20$ and $E(B-V)=0.35$, 
 adopted for foreground and cluster stars, respectively.}
\label{wfi_cmd}
\end{figure*}

The $V$ vs. \bmv~ and the $V$ vs. \vmi~ CMDs, obtained using 
the cleaned catalogs,  are shown in  
Fig. \ref{wfi_cmd}, where
horizontal bars indicate the median errors in color, while vertical bars
(barely visible),
indicate the median errors in  magnitude for one magnitude bins.
The number of stars and the error bars in the $V$ vs. \bmv~ diagram are
smaller than those in the $V$ vs. \vmi~ diagram because in the $V$ vs. \bmv~
the selection has been applied to the $B$, $V$ and $I$ magnitudes while in the 
$V$ vs. \vmi~ diagram it has been applied only
to the $V$ and $I$ magnitudes.
 
 The horizontal dashed line in each diagram,
indicating the magnitude   
completeness  limit of the \citet{sung00} catalog, gives an idea of the amount
of new photometric information yielded by this survey.

Both diagrams clearly show the upper  main sequence  
(\bmv $\lesssim 0.5$ and  \vmi $\lesssim 0.6$) dominated by the
very bright cluster stars. These are the more massive stars which, 
having a very short pre-main sequence lifetime, have already reached 
the main sequence showing a very small age spread. 

As already noted by \citet{sung00}, a sequence of stars
 appears in the range $0.7 \lesssim V-I \lesssim 1.5$
and $13 \lesssim V \lesssim 18$. These stars were interpreted by \citet{sung00}
as foreground stars less reddened than the remaining stars in the field.

Cluster stars fainter than $V\simeq 14$ are expected to be pre-main sequence
stars   which populate a large region in the CMDs,
 highly contaminated by field stars. 

One of the
 most important features of these diagrams is the well defined blue envelope;
it is due to the presence of the giant molecular cloud, which   prevents  to see
field stars (mostly main-sequence)
 more distant than the cloud. Therefore, the well
defined blue envelope of the CMDs is populated
 by main sequence field stars
   at the distance of the cloud. Background field stars, i.e.
stars more distant than the cloud, are highly obscured by the cloud and therefore
they  
  would be visible at magnitudes and colors much fainter and
much redder than their intrinsic values.

Finally, we note the presence of an unexpected  sequence of red stars, 
well separated
from the bulk of the other objects. The analysis  of the 
spatial distribution of these stars
indicates that they are uncorrelated with the region dominated by the 
cluster  stars, identified by a clustering of selected   members, as will
be shown in Sect. \ref{xwfispat_dist}. On the contrary, they cover almost
uniformly the less obscured regions of the molecular cloud and are likely
associated with clump red giants at larger distances. 
Since we are here mainly interested to study the stellar population of 
the very young cluster \ngc6530, we defer a further investigation
of these stars to a future work. 
\subsubsection{Cluster distance}
 As already mentioned in the previous section, 
   at \ngc6530 age  and distance, 
  only stars brighter than $V\sim 14 $ define the  main sequence that
 in this magnitude range is almost 
vertical   preventing  the estimate of  the cluster distance using the
usual photometric method.

Fainter cluster stars do not define any sequence, but,
on the contrary, they populate a large region   of the CMD 
which cannot be used to
constrain the cluster distance. 

  Instead, 
 we derived a reliable estimate of the cluster distance by taking advantage of
the particular feature of its CMDs:   the well defined
blue envelope. It is  populated by main sequence field stars at the cloud distance,
 which is very similar to the cluster distance
 if one does the very reasonable assumption that 
the cluster lies just before the cloud. In fact, the blue envelope
fixes a   magnitude limit  within
which we see stars (either belonging to the cluster or not),
with a reddening equal or less than the cluster reddening.
In particular, stars along the blue envelope
brighter than $V\sim14$ are  cluster stars that have
reached the main sequence, while fainter stars along the blue envelope are
foreground stars up to the cloud distance.

We    superposed  the theoretical
ZAMS computed by \citet{sies00}  on  CMDs ({\it solid line} 
in Fig. \ref{wfi_cmd})
and we  found that the
best match between this curve and the blue envelope of the star distribution
is found if  a distance of $d\simeq1250$\,pc is adopted,
corresponding to a distance modulus $(V-M_V)_0\simeq10.48$, and a reddening
$E(B-V)=0.20$ (corresponding to $A_V=0.6$ and $E(V-I)=0.26$ using the reddening
law $E(V-I)=1.3\times E(B-V)$ by \citet*{muna96}). 
We note that the  reddening $E(B-V)=0.20$ is the value
derived for foreground field stars, which suffer from a reddening significantly
smaller than  cluster stars. In fact,
for $V<14$, the theoretical model
is bluer than the bright stars 
 because these are main sequence cluster  stars at
  the same distance ($d \simeq 1250$\,pc)
   but with reddening $E(B-V)=0.35$ (corresponding to $A_V=1.1$ and 
$E(V-I)=0.46$), that is the average cluster
 reddening given by \citet{sung00},  as
 shown by the {\it thick long dashed line} in Fig. \ref{wfi_cmd}.
 For comparison   the 
ZAMS at the distance of 1800  pc, i.e. 
the distance adopted by \citet{sung00}, is shown as  {\it dashed line};
the extinction vectors $A_{V}^{\small \rm fg}=0.6$ and $A_V=1.1$, corresponding 
 to foreground stars and cluster stars respectively, are also shown
 in Fig. \ref{wfi_cmd}.  

 The distance value  $d \simeq 1250$\,pc  derived 
for \ngc6530 is very close to the lower limit of the   ample  range of values 
(1300--2000 pc)
photometrically determined for   bright stars in previous works; 
 it is, instead,  significantly larger
than the value obtained by \citet{lokt01} from the
Hipparcos  trigonometric parallaxes  of 7 stars 
(see Table \ref{distance_tab}).
%
\section{Optical Data and X-Ray Source Identifications}
%
As already discussed in the Introduction, X-ray observations
are very useful tracers of 
membership in  star formation regions or  
 very young clusters \citep[e.g.][]{alca95,rand95,ster95,flac00a},
 such as \ngc6530. 
Alternative membership assessment methods  are not suitable
 for \ngc6530, since
neither reliable proper motion  measures nor radial
velocities for magnitudes fainter than $V\sim14$ 
are yet available.

To trace the pre-main sequence locus of \ngc6530,
the  optical data derived
in this work were matched with   
X-ray sources published by  \citet{dami04}, although the Chandra ACIS FOV
 is about 4 times smaller than the WFI FOV. However, 
 the Chandra ACIS FOV is
centered very close to the \ngc6530 nominal center 
RA=$18^{\rm h}04^{\rm m}25^{\rm s}$, 
Dec=-24\deg22\min00\sec \citep{dami04},
 thus allowing  us to study most of the cluster stars. 

The total 
number of optical sources  falling in the Chandra ACIS 
FOV  is 8956, 
 while the \citet{dami04} catalog contains 884 X-ray sources. 
In order to estimate the number of X-ray sources that are not cluster members,
 \citet{dami04}  examined 
  an observation of the
 Galactic Plane obtained  with the same detector  and  used it as
 "control field". By analyzing the detected
 count-rate distributions of the two fields they conclude that at least
 $90\%$  of the X-ray sources are very probable cluster members.
 
To cross-correlate 
the X-ray and optical
 catalogs we have used a matching distance $d < 4 \sigma_X$ 
(where $\sigma_X$ is
the X-ray position error, see Damiani et al.\ 2004). In doing this, we
found and corrected a systematic shift between X-ray and WFI positions
of  0\Sec2 in RA, and  -0\Sec26
 in Dec.
For three X-ray sources with such a small X-ray error 
that $4 \sigma_X$
was less than 1\Sec5,  we relaxed the identification
condition to $d < $1\Sec5.  This resulted in a number of
multiple identifications, among which four turned into single identifications
by using
a reduced distance $d < 1.5 \sigma_X$.  This leaves us with 
721 single,
44 double  and 
3 triple
X-ray identifications in the optical catalog; in addition,
1 X-ray source corresponds to 4 optical identifications and
1 X-ray source corresponds to 6 optical identifications.  These 
multiple identifications are near the edge of the ACIS FOV
where the spatial resolution is much worse than in the center. 
 The total number of X-ray
sources with WFI counterpart(s) is therefore 770; 
of them only 15 X-ray identified stars 
come from the Sung et al. catalog  and are not 
in the WFI catalog. The 
total number of optical sources with an X-ray counterpart is 828.
  The agreement between X-ray
and WFI positions is excellent in most cases, with offsets below
1\sec. The final list, comprising 770 X-ray sources, with the corresponding
optical counterpart(s), is given in Table \ref{X_opt_data}\footnote{available
in electronic format at the Web page http://cdsweb.u-strasbg.fr/}, where
col.s 1 and 2 are the celestial coordinates, col. 3 indicates the parent
optical catalog, col. 4   is
 the optical identification number, 
col.s 5--10 are the $BVI$ magnitudes and their uncertainties, col. 11
is  the X-ray identification number of the \citet{dami04} catalog and, finally,
col. 12 is the
 number of optical sources corresponding to the 
X-ray detections.
\input{X_opt_data}

\begin{figure*}[!htb]
\includegraphics[width=16cm]{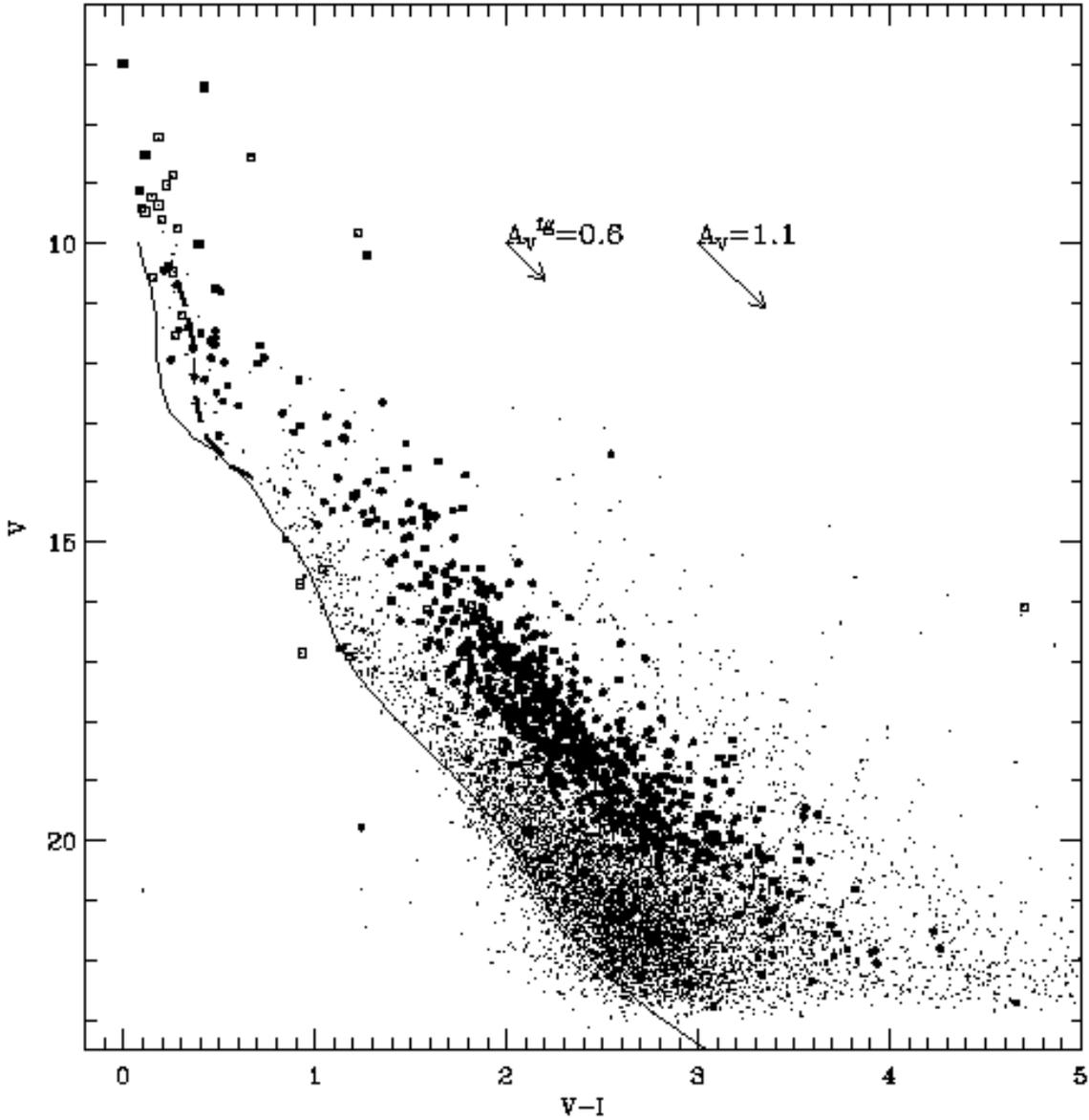}
\caption{$V$ vs. $V-I$ CMD of  stars within the
 Chandra ACIS
FOV.  {\it Dots} indicate all  stars in the optical
 catalog falling
in the ACIS FOV, while {\it large filled symbols} indicate   optical
 stars having
an X-ray counterpart;   stars added from 
  the \citet{sung00} catalog are marked by {\it squares}. 
  {\it Lines} and extinction vectors are as in
  Fig.\ref{wfi_cmd}.}
\label{opt_x_viv}
\end{figure*}

 Fig. \ref{opt_x_viv} shows
the $V$ vs. \vmi~ CMD of    
 stars falling in the Chandra ACIS FOV. 
 {\it Dots} indicate all  stars in the optical catalog falling
in the ACIS FOV, while {\it large filled symbols} indicate   optical
 stars having
an X-ray counterpart;  stars added from 
  the \citet{sung00} catalog are marked by {\it squares}.   
As expected,   X-ray detected stars trace very well the   CMD  region 
occupied by the cluster stars. The  cluster region identified
by 220 X-ray sources, at magnitudes
brighter than $V=17$, was already  analyzed in \citet{dami04},
using the photometric catalog of \citet{sung00}. With the present survey 
it is  now possible to identify the fainter cluster pre-main sequence region.
In fact, most of the stars identified with an X-ray source
are located within a well defined region,  
whose width may be due to an age spread and/or to binary stars.
 Only a small percentage of
stars with an X-ray  counterpart has  photometry that is not consistent 
with that of the cluster. The field stars 
contamination will be quantified in  
Section \ref{mass_age}, but at least qualitatively,  it  appears 
consistent with the percentage of
contaminating X-ray sources ($\lesssim 10\%$) 
given by \citet{dami04}.

As it will be shown in Section \ref{mass_age}, the cluster pre-main sequence
region   is comprised between the isochrones
of 0.3 and 10 Myr; we used these isochrones to select 
 possible cluster members  from our optical catalog.

\section{Infrared data}
\begin{figure}[!htb]
\includegraphics[width=10 cm]{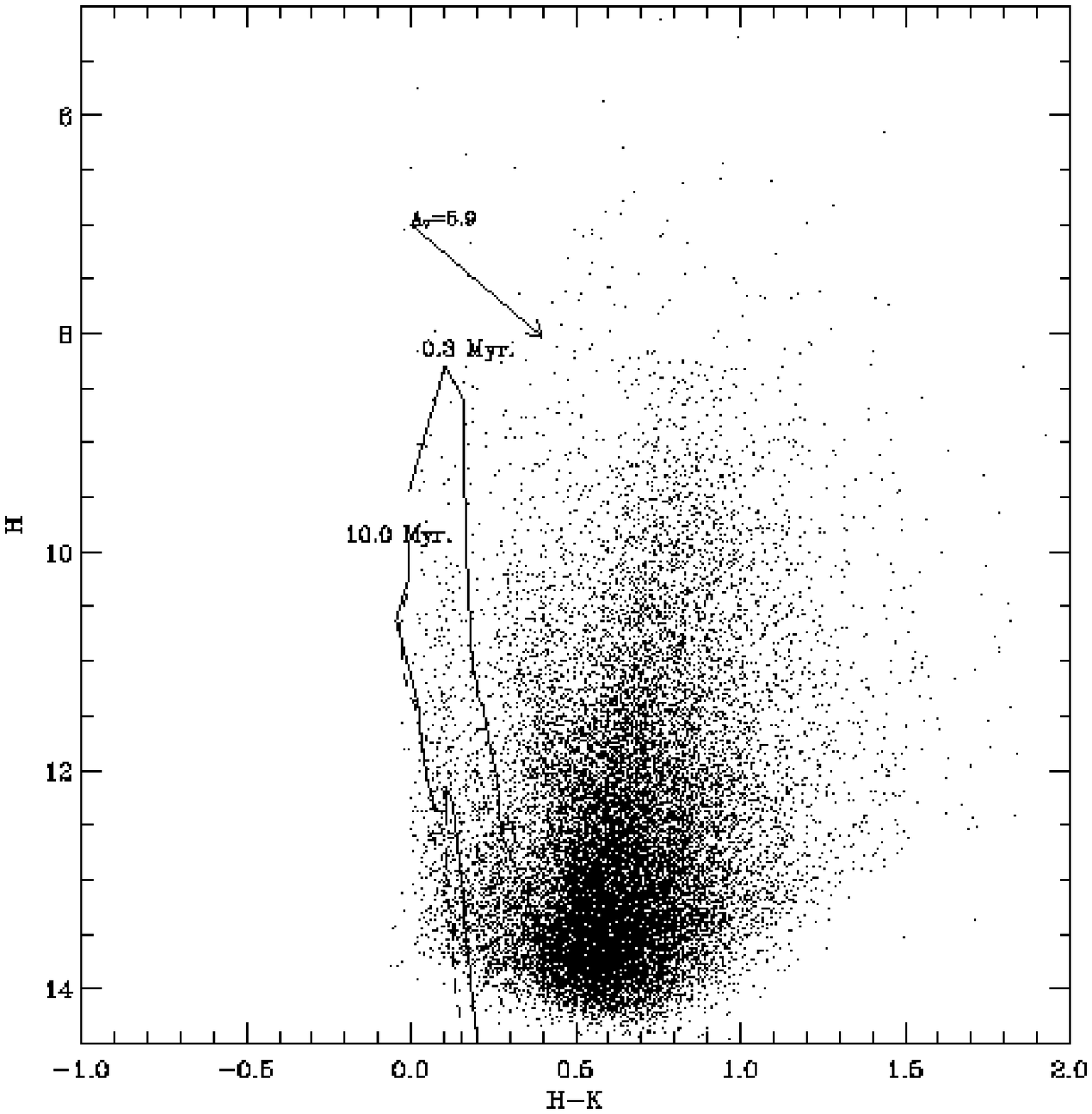}
\caption{$H$ vs. $H-K$ CMD  of all  2MASS point sources
  falling in the WFI FOV. The {\it solid lines} 
are the isochrones of 0.3 and 10.0 Myr, whereas the 
{\it dashed line} is
the ZAMS at the cluster distance $d=1250$\,pc.}
\label{wfi_hkvsh}
\end{figure}

As discussed in the previous section,   background stars are heavily obscured 
by the molecular cloud    and therefore they appear at
magnitudes much fainter than their intrinsic values.
Nevertheless, optical diagrams presented in the previous section do not allow us
 to
distinguish very faint objects with no or very low reddening, from intrinsically
bright background objects appearing as very faint stars due to their high
reddening. Since near-infrared bands   are very sensitive to reddened
stars, we  constructed the near-infrared
 CMDs using the 
$JHK_S$ magnitudes from the 2MASS  public catalog,
 with the aim to better characterize the stellar populations
 in the WFI FOV. 
The diagram   $H$ vs. $H-K$ of  stars of the 2MASS catalog falling
in the WFI FOV is
 shown in Fig. \ref{wfi_hkvsh}.
We note that only stars with  
magnitudes measured from point spread-function fitting or aperture photometry
were selected from the 2MASS catalog.

The large reddening effect can be quantified by inspecting  the  
$H$ vs. $H-K$ diagram. In fact, in this diagram most  stars are 
intrinsically colorless ($0\lesssim H-K \lesssim 0.5$, \citet{alve98})
and therefore the apparent $H-K$ color of a star is, 
to first order, a qualitative measure of the amount of
extinction toward this star. This means that, apart from relatively few stars
having  reddening of the order of the  average
 cluster reddening $E(B-V)\simeq0.35$,
corresponding to $E(H-K)\simeq0.07$,
 the bulk of the stars have colors  $H-K\gtrsim0.3$ that, using the
 \citet{math90} reddening law, corresponds to $A_V\gtrsim4.4$. 

In this diagram  we     superposed 
three theoretical curves, that are the ZAMS, the 0.3 and 10.0 Myr
 isochrones of Siess et al., 2000 (see Sect. \ref{mass_age})  
reddened using the cluster reddening value $E(B-V)$=0.35  and
 the \citet{math90} reddening law.
The location of these curves can be used to identify   
   cluster stars and   foreground field
 stars, affected by a reddening equal to or smaller than the average cluster
reddening, from the bulk of the 
  high reddened stars ($H-K\gtrsim0.3$) that are either stars
  beyond the molecular cloud  or 
  cluster members showing IR excesses, i.e. Classical
T\,Tauri stars.  
\subsubsection{Optical and 2MASS detections}
In order to better characterize the  stellar population of 
the optical catalog, we  matched optical data with the 2MASS catalog,
 using a matching radius of 0\Sec8. We found   systematic
offsets  in both right ascension and declination of -0\Sec3 and 0\Sec16, 
respectively; 
after correcting the optical catalog for this offset, a total of   
 15\,282 common stars were found. 

\begin{figure}[!htb]
\includegraphics[width=10cm]{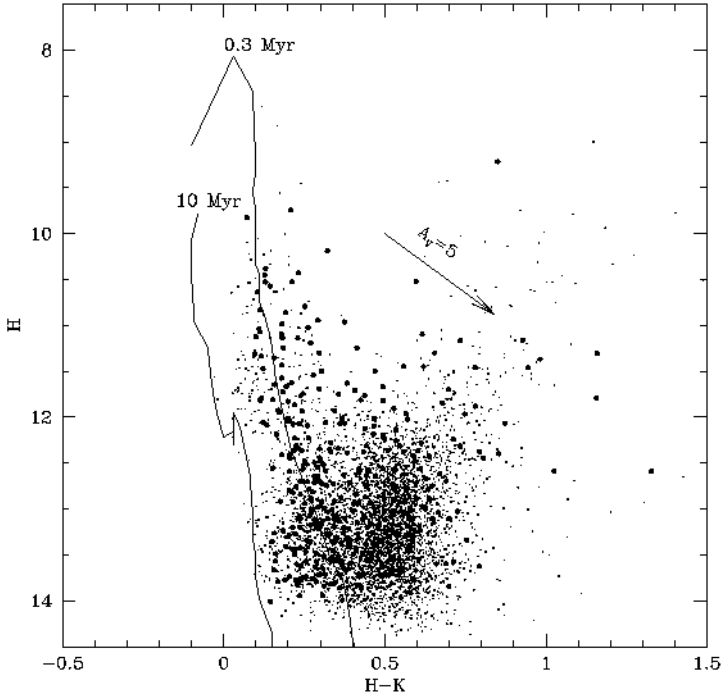}
\caption{$H$ vs. $H-K$ 
CMD  of the stars detected both in the
optical and in the near-infrared catalogs. {\it Dots} are stars  with age 
between 0.3 and 10 million years  as 
selected in the $(V-I)$ vs. $V$ diagram while {\it black bullets} are those
with an X-ray counterpart. The {\it solid lines} 
are the isochrones of 0.3 and 10.0 Myr.}
\label{2mass_wfi_x_hkvsh}
\end{figure}

\begin{figure}[!h]
\includegraphics[width=10cm]{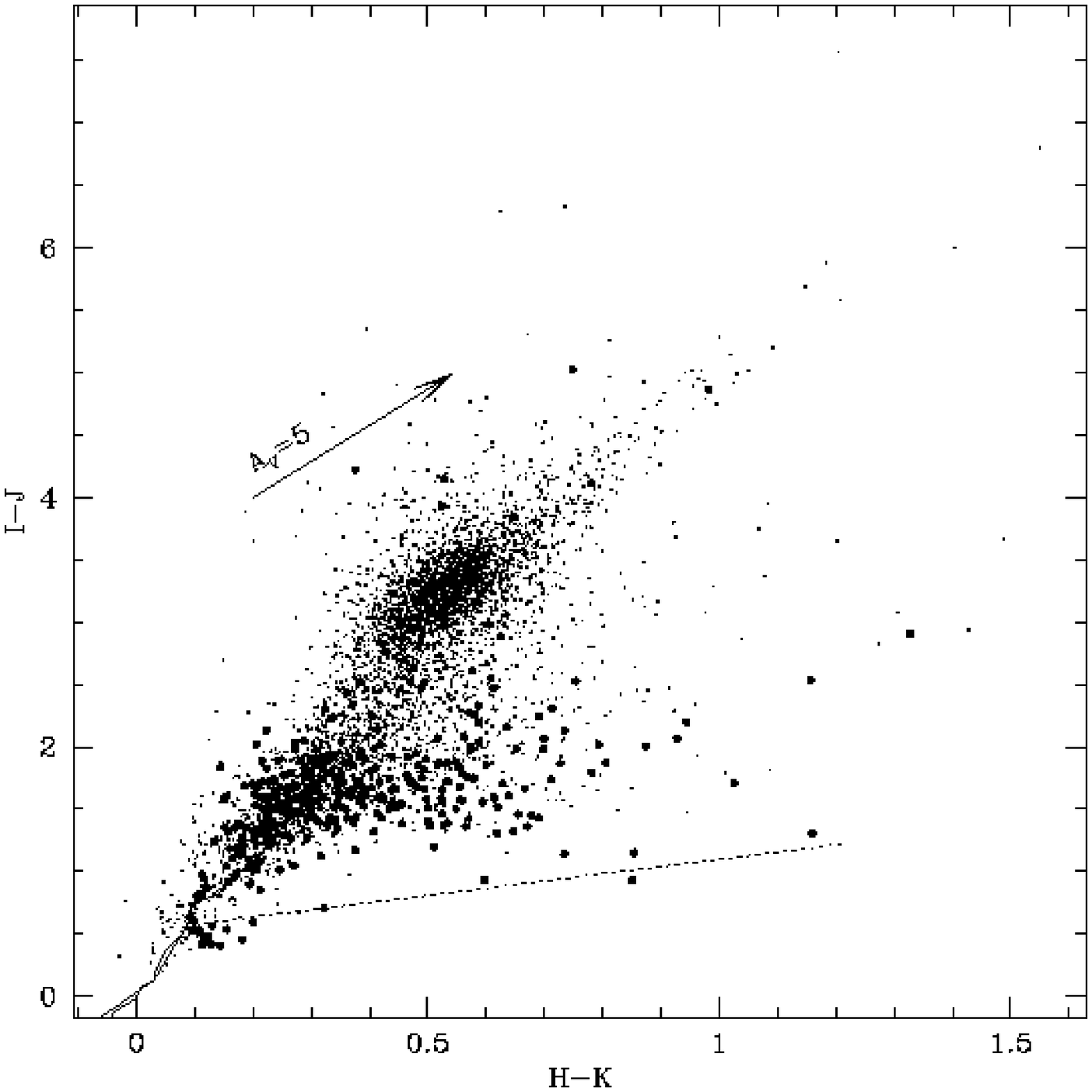}
\caption{$I-J$ vs. $H-K$ 
color-color diagram  of the stars detected both in the
optical and in the near-infrared catalogs. Symbols are as in 
Fig.\,\ref{2mass_wfi_x_hkvsh}.   The {\it solid lines} 
are the isochrones of 0.3 and 10.0 Myr, whereas 
the {\it dotted line} corresponds to the locus
of classical T Tauri stars \citep{meye97}.} 
\label{2mass_wfi_x_hkvsij}
\end{figure}

 Fig. \ref{2mass_wfi_x_hkvsh} and Fig. \ref{2mass_wfi_x_hkvsij} are 
 the
color-magnitude and the color-color diagrams 
of  possible cluster stars detected both in the
optical and in the near-infrared catalogs. They are the 9613 stars    
({\it dots})
that, in the $(V-I)$ vs. $V$ diagram,  are
found between the isochrones at
0.3 and 10.0 Myr, that is  the estimated
age spread of the cluster (see Sect. \ref{mass_age}).  Those stars
for which an X-ray counterpart was  also found are indicated as
{\it black bullets}
and, as asserted in \citet{dami04}, more than $90\%$ of them are 
cluster members.
In these diagrams,  we also plotted the isochrones of 0.3 and
10 Myr, indicated by the solid lines, and the reddening vector.
 The {\it dotted line} in Fig. \ref{2mass_wfi_x_hkvsij}
corresponds to the locus
of classical T Tauri stars \citep{meye97}.

 If we compare Fig.s \ref{wfi_hkvsh} and \ref{2mass_wfi_x_hkvsh}
we note that  most of the reddened stars (e.g. $H-K \gtrsim 0.6$) in the 
IR catalog are not seen in the optical bands, at least
 within the limiting magnitude of this survey.
Most of the stars detected both in the optical and in the near-infrared
 bands have $H-K\lesssim 0.6$, but there are objects with $(H-K)$ up to 1.5.
Using the extinction law of \citet{math90}, these values indicate that
  most of the stars in the optical catalog
have extinction   $A_V\lesssim8.8$ mag,  while the high reddening objects can have
$A_V$ up to 20 mag.

Fig. \ref{2mass_wfi_x_hkvsh} clearly shows that,  many cluster members 
({\it black bullets})
are located near the young theoretical isochrone, but a significant number
is also found redwards, with $(H-K)$ up to 1.3.
Therefore, whereas   the cluster members (selected for their X-ray emission)
 are located 
in a well defined region of the optical  CMD, they are much more "dispersed" 
in the $H$ vs. $H-K$ diagram. 
This  is  probably due to the fact 
that  a large fraction of cluster members 
 shows significant excesses in the IR colors.  

This can be   seen also in
 the $H-K$  vs. $I-J$ diagram of Fig. \ref{2mass_wfi_x_hkvsij}
where many of the 2MASS-optical candidate members are found in a strip 
around the theoretical curves, while at low reddening  
a clump of stars is found at $H-K\gtrsim0.4$ and $(I-J)\gtrsim 2.5$. 
 As already discussed,  the  latter objects are the
highest reddened stars and have $(V-I)\gtrsim 3.5$ in the $V$ vs. $(V-I)$ diagram.
Among the 2MASS-optical sources that are also X-ray detected,
less than $10\%$ are 
high reddening sources ($I-J\gtrsim 2.5$).
 Many of the X-ray detected cluster members are instead
found along the theoretical isochrones, but there is also a significant number 
of stars with large $H-K$, but in a different direction with respect to
the reddening
vector. This supports the hypothesis that these X-ray sources are
Classical T\,Tauri stars with IR excesses, as also suggested by
 \citet{dami04}.
 
These properties are found also in the other combined 
optical-IR color-magnitude and color-color
diagrams and therefore we conclude that optical and 2MASS combined data   
 alone cannot be used to select clean sample of PMS cluster members.

%
\section{Masses, Ages and Spatial Distribution}
\label{mass_age}
Using the \citet{sung00} photometric catalog,
  \citet{dami04}  determined mass and age of   X-ray detected  
stars brighter than $V=17$,
finding stars down to 0.5$M_\odot$, with  completeness   only 
for masses larger than $1.5M_\odot$. In addition, they found that the median
age of the cluster stars in the central region is 0.82\,Myr with a
maximum spread of 4\,Myr. These values were derived  assuming a cluster
  distance  larger than  the   value
estimated in this work, which  led those authors to overestimate  
luminosity and mass of the stars.

In the present work we have recomputed 
mass and age of the probable cluster members, derived by
interpolating   the  
theoretical tracks and isochrones calculated by \citet{sies00} to the
positions of the stars in the $V$ vs. \vmi~ CMD   for
which  an X-ray counterpart has been found. Of  these stars
more than $90\%$ are cluster members,  although it is unknown which
percentage they are of the whole cluster population. The completeness of this
sample is discussed in Section  \ref{lfandmf}.

We  used the  models of \citet{sies00} because it was  shown that  
masses predicted with these tracks are
consistent with  masses estimated dynamically \citep{simo00}.
In addition they are in good agreement with the pre-main sequence models of
\citet{bara98}, which are available only for ages older than 1\,Myr.
We are, however, aware of the large uncertainties of these models for age
younger than 1\,Myr, which are   based on  oversimplified initial conditions,
as discussed in \citet{bara02}.
Nevertheless, they are the most complete set of theoretical tracks
available in the literature, already used in
 very recent works on star formation regions \citep[e.g.][]{flac03a}.

Since  metallicity has never been estimated  for NGC 6530,
we assume for this cluster a solar metallicity  and we consider the
  \citet{sies00} models with $Z=0.02$, $Y=0.277$, $X=0.703$ and 
no overshooting. 
To  convert   effective temperatures and luminosities
of the adopted model
to the empirical $V$ vs. \vmi~ plane, 
we   used the 
 conversion table of \citet{keny95} where
  both optical and infrared colors are available.  

\begin{figure*}[!htb]
\includegraphics[width=16cm]{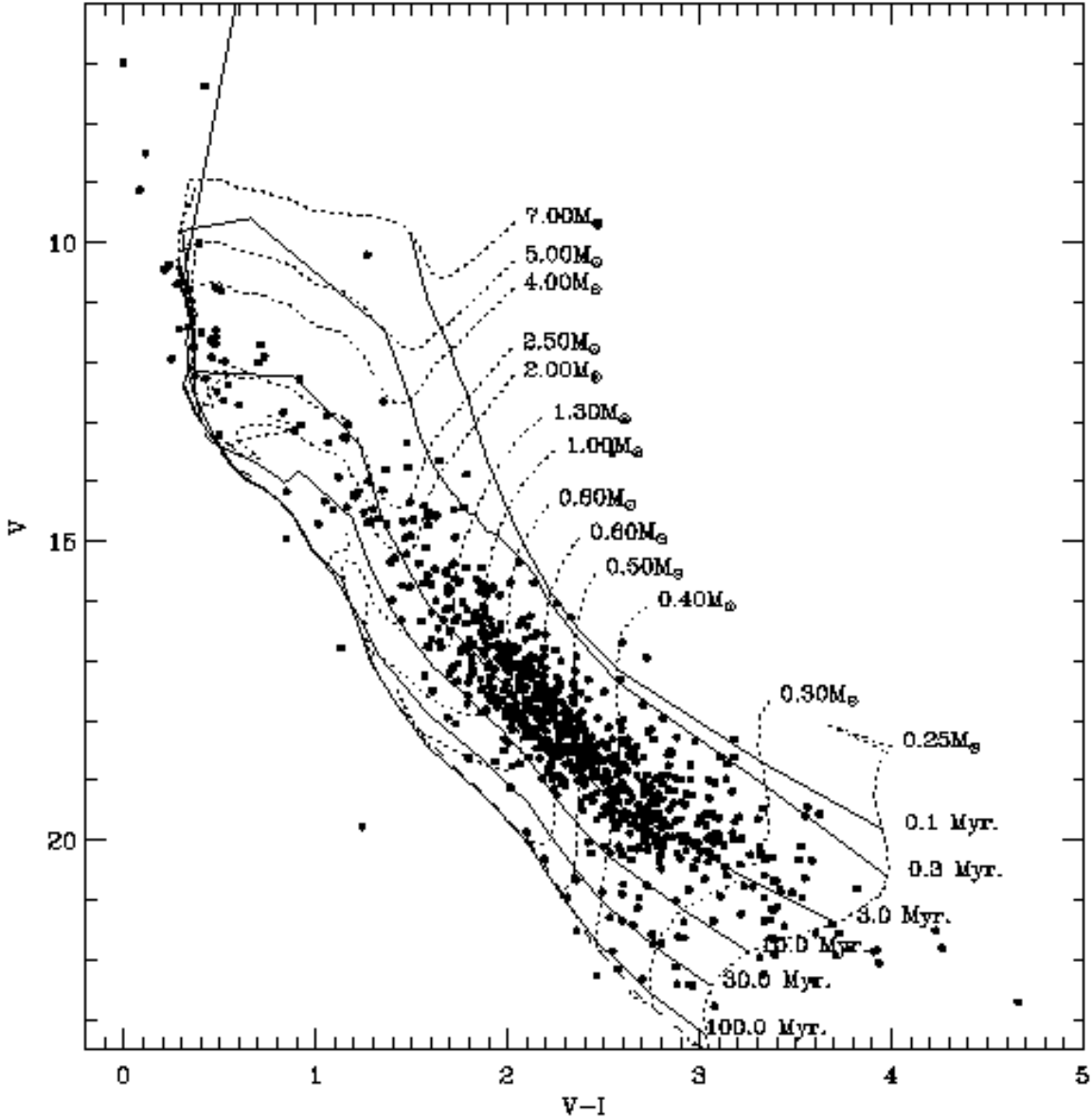}
\caption{$V$ vs. $V-I$ CMD of the stars within the 
Chandra ACIS FOV 
with an X-ray counterpart. The {\it dotted lines } are the solar
metallicity
 evolutionary tracks computed by \citet{sies00}; the corresponding
 mass value in solar mass units
is given at the redder extreme of each track.  
The {\it solid
lines} are the corresponding 
isochrones for age between 0.1 and 100 Myr, while
the {\it dashed line}
is the ZAMS at the distance of the cluster. We note that the
theoretical curves were reddened using the cluster reddening value given by
\citet{sung00} and the reddening law of \citet{muna96}.}
\label{opt_x_viv_tr_iso}
\end{figure*}

Fig. \ref{opt_x_viv_tr_iso} shows the $V$ vs. \vmi~ CMD 
  of X-ray detected cluster members 
with overimposed the adopted theoretical curves.
  We note that the curves were reddened using
the value $E(B-V)=0.35$ given by \citet{sung00} and in agreement with our 
data (see Sect. \ref{cmd_optical_data}), $A_V=3.1 \times E(B-V)$ and
  the reddening law $E(V-I)=1.3\times E(B-V)$ \citep{muna96}.  The
   absolute magnitudes of the theoretical models were
  transformed to apparent magnitudes assuming the distance
 $d \simeq 1250$\,pc, derived in Sect. \ref{cmd_optical_data}.
{\it Dotted lines} are
the evolutionary tracks for masses between 0.25 and 7 $M_\odot$,  while 
{\it solid
lines} are the isochrones for ages between 0.1 and 100 Myr, this latter
being very similar to the ZAMS  
({\it dashed line}). 

\begin{figure}[!htb]
\includegraphics[width=11cm]{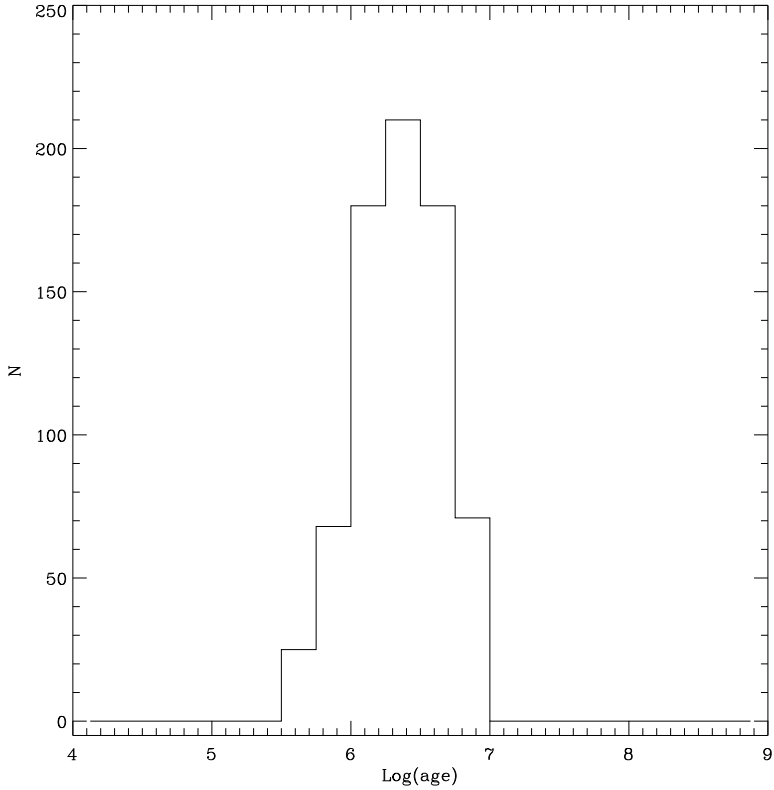}
\caption{Age distribution of all  optical/X-ray detected sources; the star
age  was  
estimated by comparing the star positions in the $V$ vs. $V-I$ CMD,
 with the isochrones computed by \citet{sies00}. 
 The median age value is 2.3 Myr.}
\label{age_hist}
\end{figure}

Due to the strong variation of the surface spatial density 
of this area,  the cluster is  expected to be 
affected by a non-negligible differential reddening.  
Evidences of such effect were found by \citet{sung00}.
This implies that the age and mass estimates are affected by
uncertainties  due to  differential reddening. However, 
as discussed in \citet{dami04}, for most of the low mass 
cluster members, the reddening vector
direction is almost parallel to isochrones and thus the age estimates are  
less affected than mass estimates that can have an uncertainty up to 0.5
$M_\odot$. Nevertheless,  this uncertainty does not drastically affect 
the determination of the Initial Mass Function  since it 
is smaller than the considered mass bins.
 Fig. \ref{opt_x_viv_tr_iso} shows that most of the X-ray detected stars 
are included between the theoretical isochrones of 0.3 and 10 Myr. We expect
that stars outside this age range are mainly contaminating objects
or highly reddened cluster members. 
We note that the observed spread (1.5--2 mag) 
 is larger than that expected from binarity
or  photometric uncertainty  since only data with errors smaller than 0.2 mag
have been  considered.
 The age  distribution for the stars in this age range
is shown in Fig. \ref{age_hist}; it is quite symmetric and  
concentrated, with a median age of 2.3\,Myr.  

\begin{figure*}[!htb]
\includegraphics[width=14cm]{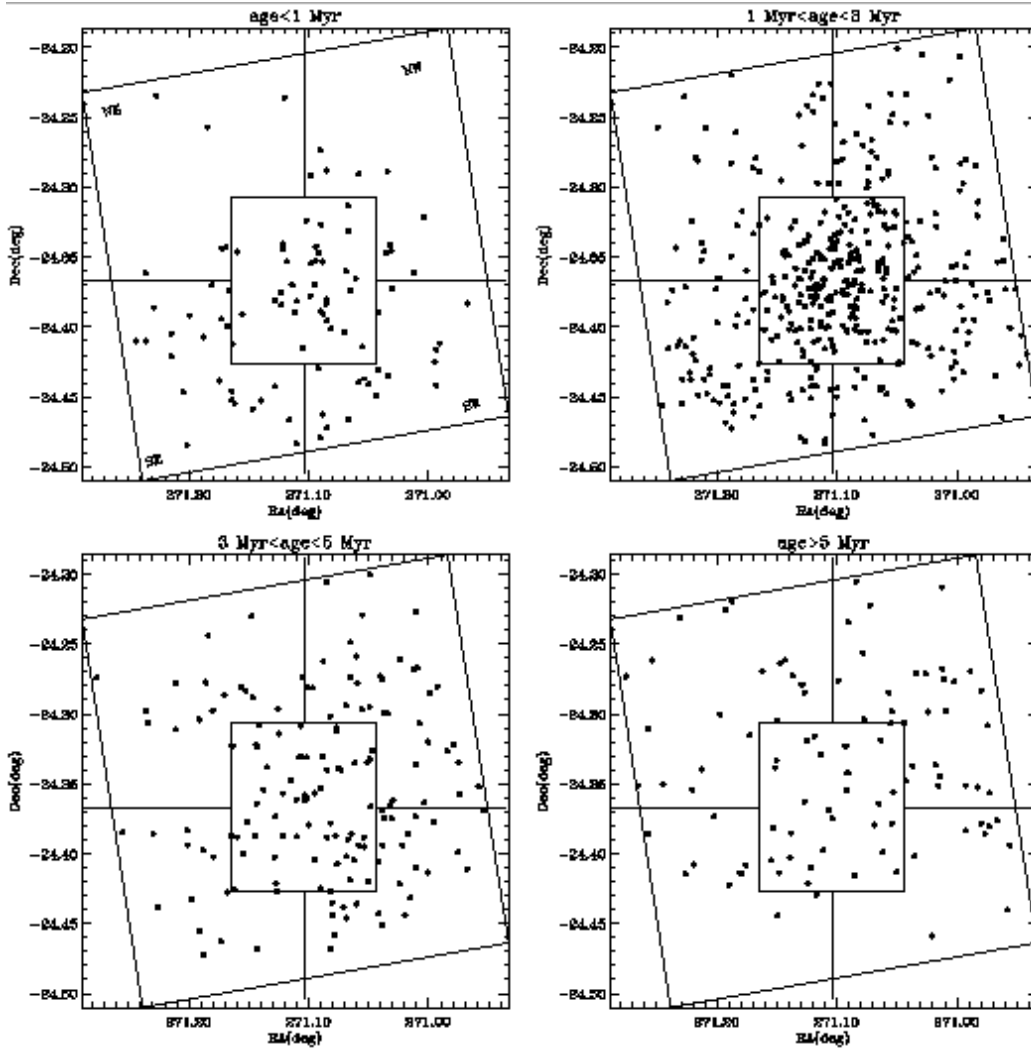}
\caption{Spatial distributions of the optical/X-ray detected stars in the four
age ranges.}
\label{radec_bin_age}
\end{figure*}

\begin{figure}[!htb]
\includegraphics[width=10cm]{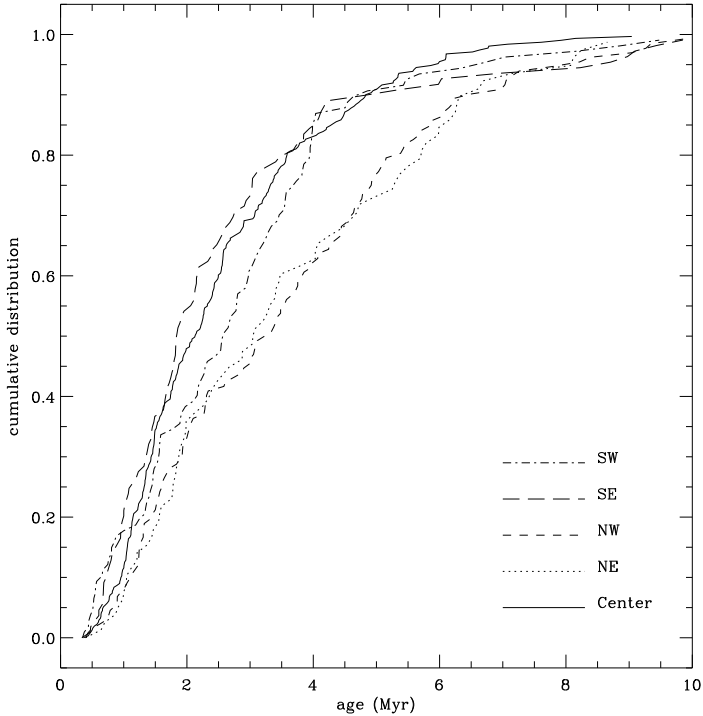}
\caption{Age distributions of the optical/X-ray sources in  the five
distinct spatial subregions defined in Fig. \ref{radec_bin_age}.}
\label{cum_age}
\end{figure}

Due to the position of the cluster with respect to the Hourglass nebula, it was
proposed for this cluster a sequential process of the star formation
\citep{lada76}. In
particular, \citet{dami04} suggested  that the star formation has progressed
from north to south. In order to find further evidences of this effect, we  
plotted in Fig. \ref{radec_bin_age} the spatial distributions of the
optical/X-ray detected sources 
in   four different age ranges. Our
results 
 support the conclusion of \citet{dami04} since very young stars are present
(almost) only in the cluster
  southern region, while the spatial distribution become more
uniform for older ages. 
To test the validity of this conclusion, we have done a 
statistical analysis
using the age distributions in 5 different spatial 
regions, namely the cluster center
included in a square region of side 7.2 arcmin, as defined in \citet{dami04},
and 4 outer regions (see Fig. \ref{radec_bin_age}).
 The cumulative  age distributions  
  in Fig.\,\ref{cum_age} show a clear age difference
 between the southern and central regions and the northern ones. Using the
 two sample Kolmogorov-Smirnov tests, we find that 
 the   cumulative distribution functions of each of the southern and central
  regions are 
  different  from  the northern regions at
  a significance level greater than $99\%$.
 
\begin{figure*}[!htb]
\includegraphics[width=14cm]{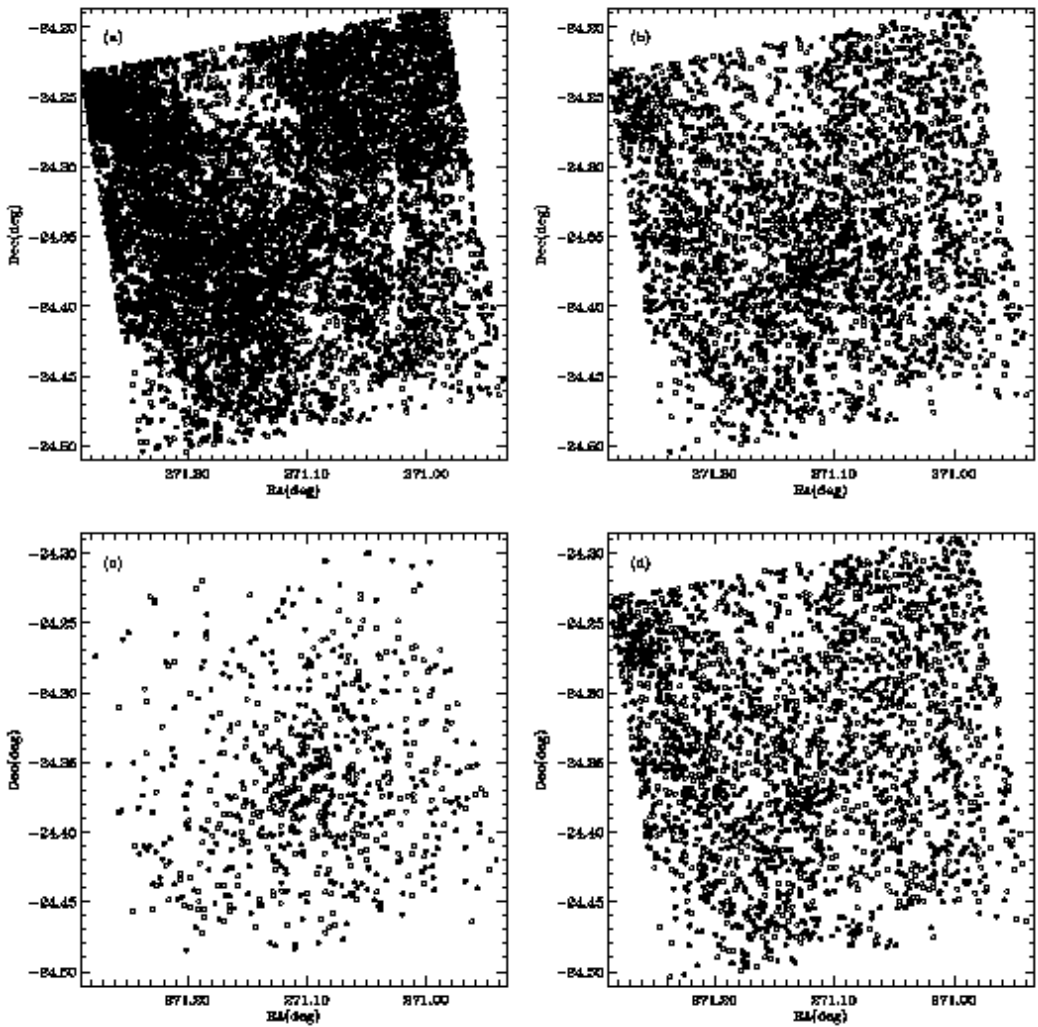}
\caption{
({\it a}): spatial distribution of all the 8975
optical sources  within the Chandra ACIS FOV; 
({\it b}): same distribution for the 3427 stars with
magnitudes and colors consistent with that of the cluster, whose
age is between 0.3 and 10.0 Myr
(referred to as
"optical candidate members"); 
({\it c}): spatial distribution of the 737 Optical/X-ray cluster
members; 
({\it d}): spatial distribution of the 2690 optical 
candidate member sample after subtracting the 737 optical/X-ray cluster members.}
\label{radec_membership}
\end{figure*}

Fig. \ref{radec_membership}{\it a} 
shows the spatial distribution of all the 8975 optical
sources in the Chandra ACIS FOV to which
no photometric selection was  applied. This distribution
is clearly dominated by field stars and   does not show 
any evidence of a central clustering. 
The very strong absorption of the nebula 
causes a very patchy spatial distribution of  stars.
 Fig. \ref{opt_x_viv_tr_iso}  
shows that most of the optical/X-ray 
sources have age in the range 
(0.3--10.0) Myr.
 This means that stars with age outside
this  range are not photometrically cluster members. Therefore, we 
selected all optical stars with age in this range and we  defined them
as "optical candidate cluster members". 
The spatial distribution of these stars is shown
in Fig. \ref {radec_membership}{\it b}. 
Although it is still
dominated by field stars, some evidence of star clustering  is visible.

From this sample, we  defined as "optical/X-ray cluster members" all the 737 
optical sources with 
an X-ray counterpart and age consistent with that of the cluster.
The spatial distribution of these stars is shown in
 Fig. \ref {radec_membership}{\it c}, where a
strong evidence of clustering is seen. 
 The clustered distribution of these stars 
allows us 
to confirm the conclusion that most of the optical/X-ray detected stars 
are cluster members, and clearly shows how X-ray
observations are efficient to select cluster members.
We note that these latter stars are about $90\%$ of the whole sample of
optical/X-ray sources (828 stars), 
in agreement with the field star contamination
estimated in \citet{dami04}.

Fig. \ref {radec_membership}{\it d}
shows the spatial distribution of  the optical 
candidate cluster sample from which
the 737 optical/X-ray cluster members were
subtracted. These stars do not show any clustering evidence; on the contrary
they show  the same patchy pattern of the contaminated sample and
this suggests
that most of them are field stars.  We conclude that our optical/X-ray sample
suffers from a small contamination, and includes a very large fraction of the
cluster members within the studied magnitude limit.
\label{xwfispat_dist}
\section{Luminosity and Initial Mass Functions }
\label{lfandmf}
\begin{figure}[!htb]
\includegraphics[width=10cm]{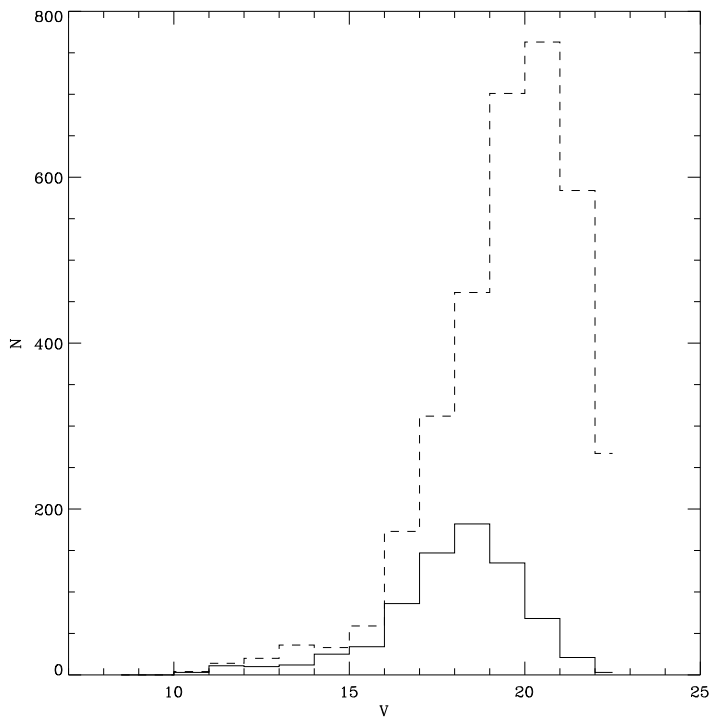}
\caption{$V$ magnitude distribution  of the  X-ray cluster members 
({\it solid histogram}) and  of  all optical 
candidate cluster members
({\it dashed histogram}).}
\label{lf_memb}
\end{figure}

Fig. \ref{lf_memb} 
shows the $V$ magnitude distribution  of the   optical/X-ray cluster members
({\it solid histogram}) and that of all  optical 
candidate cluster members
({\it dashed histogram}) defined in Sect. \ref{mass_age}. 
These distributions can be considered only as 
lower and upper  limits to the Luminosity Function of the cluster.   
In fact, due to the limited sensitivity of the Chandra ACIS detector,
we expect that 
the sample of 
"optical/X-ray cluster members" is not complete  but, how we have discussed 
above, the expected incompleteness should be small notwithstanding the sample
includes only about $21\%$ 
of the  "optical candidate cluster members". 
Due to the strong gradient of the spatial density
distribution   of the whole region, it is impossible
 to choose a suited 
 region  of the FOV  to be used as "control field" to
estimate the field star contamination of the optical 
candidate cluster members. 

 In addition, in the case of pre-main sequence stars, 
the observed luminosity function cannot be directly converted into mass
function.
For these reasons, we have constructed the mass function using  
the "incomplete"
sample of "optical/X-ray cluster members", that includes stars 
with age derived from the CMD  between 0.3 and 10 Myr. 
For these stars, we directly estimated   
 the stellar masses from their position on the $V$ vs. $V-I$
 CMD, as described in Sect. \ref{mass_age}. 
In order to obtain the mass function   of the
whole cluster, we  first considered the completeness
of the optical catalog, using the results of the artificial star tests, 
described in Section \ref{cmd_optical_data}. To obtain the optical complete
 mass distribution, 
 the fractions  of the 
retrieved artificial stars in each  $V$ and $I$ magnitude bin,
were interpolated to the position of the stars  
in the $V$ vs. \vmi~ CMD. 
The minimum  fraction computed in each mass bin is reported in
col. 2 of Table \ref{xcorr}.

To take into account  the X-ray incompleteness of member
selection, we  
corrected the mass function
assuming that the fraction of \ngc6530 members
detected as X-ray sources is, at the same sensitivity, equal to  that
detected  for a cluster of similar age for stars of a given mass.

The most suitable  region for this comparison is certainly  
the Orion Nebula Cluster, one of the best studied clusters,
for which all cluster members are known from  extensive 
optical and infrared surveys
and for which Chandra   X-ray observations were 
recently published by \citet{flac03a} and \citet{feig02}. Fig. 3 of  
 \citet{flac03a} shows the cumulative X-ray Luminosity Functions relative to
the complete optical sample of stars in the Orion Nebula Cluster,
for eight ranges of mass from 0.1 to 50 $M_\odot$. In order to obtain an
appropriate 
correction to the \ngc6530  Mass Function derived from the optical/X-ray
star sample, we  considered that the sensitivity of the \ngc6530
Chandra ACIS observation  allows us to measure X-ray luminosities
greater than  ${\rm Log}(L_X)\simeq29.2$(erg/sec) in the central region and 
greater than
${\rm Log}(L_X)\simeq29.7$(erg/sec) in the external region of the ACIS FOV
 for an X-ray spectrum with {\it k}T=1keV, and $ N_H$ corresponding to
the optical extinction (Damiani et al., in preparation). 
For each mass range considered by \citet{flac03a}, we took the fraction  of 
X-ray detected stars above ${\rm Log}(L_X)=29.2$\,(erg/sec) and
 ${\rm Log}(L_X)=29.7$\,(erg/sec), given in  columns 3 and 4 of
 Table \ref{xcorr}, and we
  used these fractions to obtain the minimum and  maximum correction to our
incomplete Mass Function. 
\input{mf6530}
 The   mass function of \ngc6530 $\Delta N/\Delta log M$ is therefore 
taken as the average of the minimum and  maximum distributions 
$(\Delta N/\Delta log M)_{\rm min}$ and $(\Delta N/\Delta log M)_{\rm max}$,
 resulting from these corrections.  
\begin{figure}[!htb]
\includegraphics[width=10cm]{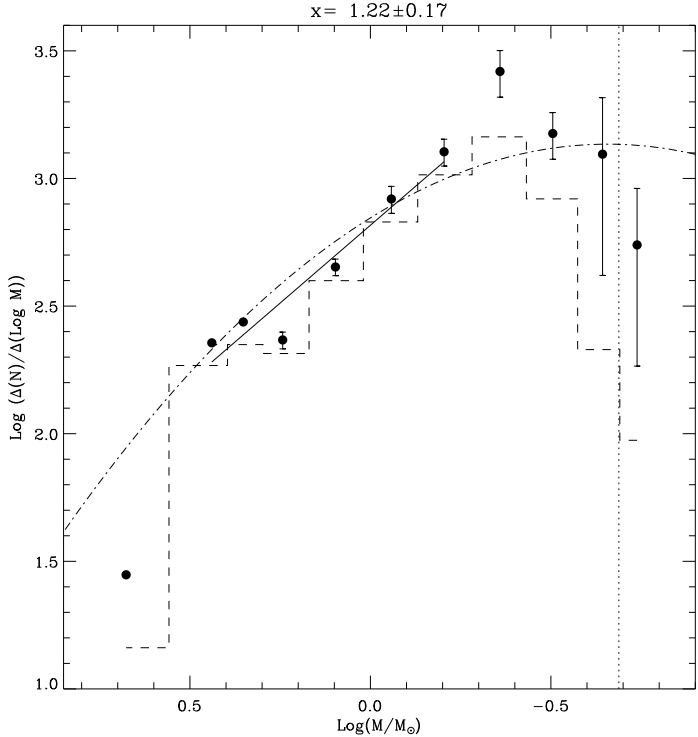}
\caption{The {\it dashed histogram} shows the X-ray incomplete
 mass function derived from the X-ray
cluster members, while the {\it black bullets} indicate  the 
mass function of the cluster corrected according to the procedure described in
Sect. \ref{lfandmf} (average 
values between the minimum and maximum corrections,
indicated in  figure by the error bars). 
  The field star mass function given
in Chabrier (2003),  arbitrarily normalized to the present data, is plotted
as the {\it dotted-dashed line}.
The {\it solid line} is the power law fitting 
the \ngc6530 mass function in the range (0.6--4.0)\,$M_\odot$.
 The power law index and the corresponding rms  uncertainty are also indicated.
The vertical {\it dotted line} corresponds to the value at which the 
photometric  completeness limit is larger than
40\%.}
\label{mf6530}
\end{figure}

The results are given in Table \ref{xcorr} and  shown in 
Fig. \ref{mf6530} where the {\it dashed histogram} shows the X-ray incomplete
 mass function derived from the X-ray
cluster members, while  the {\it black bullets} indicate  the 
corrected mass function of the cluster; the error bars show 
the minimum and maximum corrected distributions.
 Note  that, with the exception  of the lowest mass bin,  
 the optical incompleteness is always smaller than $60\%$ and it makes
 sense to correct it as described before.
 
Since the median age of the cluster is smaller than the typical time scale 
for  dynamical evolution ($\sim 100$\,Myr) 
and/or  stellar evolution,
 the measured mass function is equal to the Initial Mass Function. 
The cluster IMF increases going down 
from 6.5 to about 0.4\,$M_\odot$, where it shows a
  peak, and  decreases for lower masses. 
 In the mass range (0.6--4.0)\,$M_\odot$, it was  fitted with a power 
  law of index $x=1.22\pm0.17$ that is consistent with the  
  Salpeter index  1.35.  The lower limit of the mass range was chosen equal to
  that used by \citet{muen02a} in Orion Nebula cluster,
   in order to compare the slope of the IMF of
  these similar clusters in the same mass range.
 The  power law index obtained for \ngc6530
   is also consistent with the   index obtained for the coeval 
  Trapezium, Taurus and IC 348 clusters in the same mass range
   \citep{muen02b,muen02a,bric02,muen03}.
 In the common mass range,
  while the IMF of most star forming regions and open clusters, such as
  Trapezium, IC 348, Lambda Orionis \citep{barr04}, and Pleiades \citep{bouv98} 
      flattens at about 0.8\,$M_\odot$, the   \ngc 6530 IMF appears to
      decrease for mass as lower than about 0.4\,$M_\odot$.
    A quantitative comparison with the IMF of other star forming regions
   or open clusters for very low masses
  cannot be performed with available data, 
  since the \ngc 6530 IMF is limited to
    $M \ge 0.2\,M_\odot$.
  
  The \ngc6530 IMF has also been compared to the   field
 star IMF derived by \citet{chab03a} that is displayed in Fig. \ref{mf6530}
 by the {\it dotted-dashed line}. The  \ngc6530 IMF is  reminiscent of
  the lognormal shape derived for the galactic disk field stars.
 
 The total mass of the optical/X-ray  members is about 560\,$M_\odot$; 
 using the corrections for optical and X-ray incompleteness
 we  find that the total mass of the cluster,  including stars down to
 0.4\,$M_\odot$, is
 between 700 and 930\,$M_\odot$. 
\section{Summary and Conclusions}
We  used {\em BVI} images taken with the  WFI camera of the
 ESO/2.2\,m, available at the 
ESO/ST-ECF Science Archive, to obtain multi-band photometry of the very young
open cluster \ngc6530. This cluster is
 located in a star formation region   known as
 Lagoon Nebula (M8) that is one of the brightest nebulae in the solar vicinity.
The whole field shows  evidences of strong obscuration due to the dense
molecular cloud located just behind the cluster. 

The present photometric catalog reaches down to $V\simeq23$ and allows us to
significantly increase the knowledge of this cluster. In fact,  
the most recent photometric survey of this cluster  was published by
\citet{sung00} and reaches only down to $V=17$. From   their
survey  \citet{sung00} derived a cluster distance of  1800 pc 
corresponding to a distance modulus of $(V-M_V)_0=11.25$,
 which is much larger than the value d=1250\,pc, inferred from
the present work by considering a more complete CMD and  taking advantage
of the total absence of background stars beyond the   main sequence locus
at the cluster distance. 

The near-infrared CMDs of the same FOV, obtained
using {\em JHK} magnitudes taken from the 2MASS
catalog, suggest that this region includes a large number of
 very reddened objects that
are not present in the optical CMDs, at least within our
limiting magnitude. Few of them have been detected as X-ray sources.

Optical  and  near-infrared data  alone do not allow us to identify the
cluster stars in the CMDs, since they do not form a well
defined sequence but, on the contrary, they populate a wide region in the
CMDs. For this  reason it has been  crucial to
cross-correlate optical data with X-ray sources detected with Chandra ACIS
instrument and very recently published by \citet{dami04}. A total of 
 828  optical/X-ray sources were found and placed  in the CMD.
  $90\%$ of these stars are very probable cluster members and 
identify a very well defined pre-main
sequence region, that is the cluster locus  comprised between the
0.3 and 10 Myr isochrones. 

Masses and age of these cluster members were estimated using 
the evolutionary tracks computed by 
\citet{sies00}.  The age distribution of these stars
indicates that the median age of the cluster is about 
 2.3\,Myr . 
  A statistically significant
  trend in the spatial distribution of these stars was
  found as a function of   age, as already suggested by \citet{lada76}
  and \citet{dami04}.
This suggests a sequential star formation process from north to south.

The Initial Mass Function derived from the  photometrically selected
stars with an X-ray counterpart
 was   corrected  for incompleteness of X-ray data 
assuming that the fraction 
of \ngc6530 members
detected as  X-ray sources is, at each given sensitivity, the same of that
detected in the Orion Nebula Cluster.   In the mass range 
(0.6--4.0)\,$M_\odot$, 
the  corrected Initial Mass 
Function can be represented by a power law with 
 index $x=1.22\pm0.17$, consistent with the Salpeter index 1.35,
while at smaller masses it shows a peak and then it starts to
  decrease. The resulting IMF is  similar to that obtained for   coeval 
 clusters, such as Trapezium, Taurus and IC348,
  in the same mass range.
\begin{acknowledgements}
 This work is part of the PhD thesis of L.P.; we acknowledge  financial support
 from italian MIUR, and E. Flaccomio for useful suggestions that greatly 
 improved our analysis. We thank  an anonymous referee for useful comments and
 suggestions.  
\end{acknowledgements}
\bibliographystyle{aa}
\bibliography{biblio}
\end{document}

%% file: wfi_sung_wk.tex
\begin{deluxetable}{cccccccccccccccccccc}
\tabletypesize{\tiny}
\tabcolsep 0.1truecm
\rotate
\tablecaption{Cross-identifications of this catalog with the previous
works of \citet{walk57,kila77} and \citet{sung00} \label{wfi_sung_wk}}
\tablewidth{0pt}
\tablehead{\colhead{ID}&
\colhead{RA(2000)}&\colhead{Dec(2000)}&
\colhead{$B$}&\colhead{$\sigma_B$}&\colhead{$V$}&\colhead{$\sigma_V$}&
\colhead{$I$}&\colhead{$\sigma_I$}&\colhead{$ID_{SBC}$}&
\colhead{$\Delta$ RA\tablenotemark{a}}&
\colhead{$\Delta$ Dec\tablenotemark{a}}&
\colhead{$\Delta V$\tablenotemark{a}}&
\colhead{$\Delta (B-V)$\tablenotemark{a}}&
\colhead{$\Delta (V-I)$\tablenotemark{a}}&
\colhead{$ID_{\rm WK}$}&
\colhead{$\Delta$ RA\tablenotemark{b}}&
\colhead{$\Delta$ Dec\tablenotemark{b}}&
\colhead{$\Delta V$\tablenotemark{b}}&
\colhead{$\Delta (B-V)$\tablenotemark{b}} \\
\colhead{} & \colhead{(h m s)} & \colhead{(d m s)} & \colhead{}&
\colhead{} & \colhead{} & \colhead{} & \colhead{}&
\colhead{} & \colhead{} & \colhead{(arcsec)} & \colhead{(arcsec)}&
\colhead{} & \colhead{} & \colhead{} & \colhead{}&
\colhead{(arcsec)} & \colhead{(arcsec)} & \colhead{} & \colhead{}\\}
\startdata
WFI   5173& 18 04 35.45&  -24 32 28.82& 14.749&  0.002& 13.889&  0.003& 12.860&  0.004&   732&     0.29982&    -0.36392& -0.061&  0.016&  0.022& &  &  & &\\
WFI   5249& 18 04 45.36&  -24 32 26.44& 17.692&  0.005& 16.694&  0.003& 15.477&  0.006&   836&     0.39976&    -0.69351& -0.028& -0.040&  0.006& &  &  & &\\
WFI   5377& 18 04 33.15&  -24 32 22.37& 18.709&  0.004& 16.601&  0.003& 14.008&  0.003&   712&     0.39976&    -0.39825& -0.036&  0.177&  0.072& &  &  & &\\
WFI   5387& 18 04 13.26&  -24 32 22.09& 15.973&  0.006& 15.115&  0.006& 14.091&  0.003&   451&     0.49970&    -0.34332& -0.108& -0.007&  0.009& &  &  & &\\
WFI   5449& 18 04 31.27&  -24 32 20.17& 13.460&  0.008& 13.049&  0.009& 12.605&  0.012&   687&     0.69958&    -0.32959& -0.038&  0.069&  0.030&   131&     3.69780&    -2.16980& -0.031 &  0.051\\
WFI   5472& 18 03 47.86&  -24 32 19.58& 17.853&  0.003& 16.789&  0.003& 15.587&  0.006&   207&     0.29982&    -0.52872& -0.080&  0.028&  0.024& &  &  & &\\
WFI   5483& 18 04 37.62&  -24 32 19.21& 15.493&  0.023& 14.305&  0.012& 12.811&  0.031&   756&    -0.19988&    -0.26779&  0.001&  0.181& -0.021& &  &  & &\\
WFI   5493& 18 03 51.58&  -24 32 18.72& 15.619&  0.003& 14.779&  0.001& 13.778&  0.004&   247&    -0.69958&    -0.46005& -0.061&  0.035&  0.024& &  &  & &\\
WFI   5556& 18 03 48.69&  -24 32 16.89& 15.556&  0.002& 14.596&  0.003& 13.327&  0.005&   218&    -0.49971&    -0.53558& -0.107&  0.124&  0.030& &  &  & &\\
WFI   5660& 18 03 48.32&  -24 32 13.42& 14.556&  0.005& 13.645&  0.003& 12.635&  0.010&   212&    -0.19988&    -0.44632& -0.069&  0.033&  0.017& &  &  & &\\
WFI   5828& 18 03 43.17&  -24 32  7.96& 18.018&  0.003& 16.018&  0.005& 13.681&  0.006&   168&    -0.69960&    -0.50125& -0.031&  0.129&  0.102& &  &  & &\\
WFI   5857& 18 03 36.99&  -24 32  7.15& 17.187&  0.003& 14.814&  0.004& 11.861&  0.010&   117&    -0.49972&    -0.47379& -0.005&  0.143&  0.129& &  &  & &\\
WFI   5875& 18 03 50.34&  -24 32  6.68& 19.092&  0.007& 16.951&  0.005& 14.451&  0.003&   234&     0.00000&    -0.51498&  0.021&  0.091&  0.114& &  &  & &\\
WFI   5965& 18 03 45.52&  -24 32  3.92& 18.660&  0.004& 16.605&  0.003& 14.090&  0.005&   187&    -0.09994&    -0.45319& -0.049&  0.128&  0.083& &  &  & &\\
WFI   5979& 18 03 30.39&  -24 32  3.45& 17.355&  0.011& 16.313&  0.007& 15.024&  0.016&    83&    -0.59966&    -0.70038& -0.141&  0.045&  0.031& &  &  & &\\
WFI   5995& 18 03 51.04&  -24 32  3.01& 17.879&  0.005& 15.416&  0.002& 12.456&  0.013&   241&     0.19989&    -0.44632&  0.028&  0.122&  0.102& &  &  & &\\
WFI   6004& 18 03 59.02&  -24 32  2.75& 16.633&  0.002& 15.602&  0.002& 14.406&  0.003&   303&     0.00000&    -0.40512& -0.034& -0.027&  0.037& &  &  & &\\
WFI   6057& 18 04 29.68&  -24 32  0.99& 19.260&  0.006& 16.643&  0.004& 13.365&  0.004&   675&     0.79956&    -0.43945&  0.071&  0.072&  0.140& &  &  & &\\
WFI   6149& 18 04 20.02&  -24 31 57.73& 17.911&  0.002& 16.289&  0.006& 14.262&  0.002&   541&    -0.29984&    -0.45319& -0.005&  0.004&  0.058& &  &  & &\\
WFI   6278& 18 03 43.40&  -24 31 52.64& 18.538&  0.003& 16.515&  0.004& 14.071&  0.005&   172&    -0.19989&    -0.47379& -0.049&  0.064&  0.077& &  &  & &\\
\enddata
\tablenotetext{a}{SCB comparison}
\tablenotetext{b}{Walker-Kilambi comparison}
\tablecomments{The complete table is available in electronic
 format at the Web page http://cdsweb.u-strasbg.fr/}
\end{deluxetable}

%% file: X_opt_data.tex
\begin{table*}
\centering
\tabcolsep 0.1truecm
\caption{Cross-identifications of the optical catalog
with the X-ray source catalog of Damiani et al. (2004).}
\vspace{0.5cm}
\begin{tabular}{cccccccccccc}
\hline
\hline
\multicolumn{1}{c}{RA(2000)}&
\multicolumn{1}{c}{Dec(2000)}&
\multicolumn{1}{c}{Cat}&
\multicolumn{1}{c}{ID}&
\multicolumn{1}{c}{$B$}&
\multicolumn{1}{c}{$\sigma_B$}&
\multicolumn{1}{c}{$V$}&
\multicolumn{1}{c}{$\sigma_V$}&
\multicolumn{1}{c}{$I$}&
\multicolumn{1}{c}{$\sigma_I$}&
\multicolumn{1}{c}{ID(X)}&
\multicolumn{1}{c}{Mult.} \\
(h m s)    & (d m s)      &    &        &       &       &       &       &       &       &      &\# \\ 
\hline\\
 18 4 48.38&  -24 29  3.94& WFI&   10889& 17.346&  0.008& 15.901&  0.008& 14.011&  0.005&   811&  1\\
 18 4 26.49&  -24 29  0.30& WFI&   10947& 20.717&  0.109& 18.975&  0.024& 16.096&  0.031&   528&  2\\
 18 4 26.46&  -24 28 59.95& WFI&   10956& 18.314&  0.019& 16.729&  0.011& 14.771&  0.007&   528&  2\\
 18 4 30.41&  -24 28 53.16& WFI&   11091& 20.006&  0.010& 18.289&  0.018& 15.930&  0.004&   609&  1\\
 18 4 26.29&  -24 28 48.61& WFI&   11173& 20.225&  0.013& 18.809&  0.011& 16.370&  0.004&   524&  1\\
 18 4 21.80&  -24 28 46.25& WFI&   11216& 22.255&  0.065& 20.345&  0.024& 16.760&  0.007&   417&  1\\
 18 4 31.37&  -24 28 44.49& WFI&   11246& 21.783&  0.046& 19.634&  0.034& 16.536&  0.004&   628&  1\\
 18 4 22.54&  -24 28 40.75& WFI&   11308& 15.464&  0.003& 14.705&  0.004& 13.693&  0.005&   432&  1\\
 18 4 16.76&  -24 28 36.94& WFI&   11376& 11.803&  0.038& 11.580&  0.003& 11.102&  0.005&   305&  1\\
 18 4 23.34&  -24 28 31.98& WFI&   11471& 19.082&  0.018& 17.726&  0.004& 15.938&  0.004&   451&  1\\
\hline
\hline
\end{tabular}
\label{X_opt_data}
\end{table*}

%% file: mf6530.tex
\begin{deluxetable}{cccccc}
\tabcolsep 0.1truecm
\tablecaption{Mass Function of NGC 6530. \label{xcorr}}
\tablewidth{0pt}
\tablehead{\colhead{$\Delta M$}&\colhead{\scriptsize Min. optical}&
\colhead{\scriptsize X-ray fraction for}&
\colhead{$(\frac{\Delta N}{\Delta {\rm Log} M})_{\rm min}$}&
\colhead{$(\frac{\Delta N}{\Delta {\rm Log} M})$}&
\colhead{$(\frac{\Delta N}{\Delta {\rm Log} M})_{\rm max}$} \\
\colhead{$(M_\odot)$}&\colhead{\scriptsize completeness}&
\colhead{\scriptsize {\rm Log}($L_X$)=29.7 ~{\rm Log}($L_X$)=29.2}&
\colhead{}&\colhead{}&\colhead{}}
\startdata
  0.16 --  0.20&  0.1&  0.1~~~~~~~~~  0.5&      184&      549&      914\\
  0.20 --  0.25&  0.4&  0.1~~~~~~~~~  0.5&      417&     1245&     2073\\
  0.25 --  0.38&  0.9&  0.5~~~~~~~~~  0.7&     1189&     1500&     1812\\
  0.38 --  0.50&  0.9&  0.5~~~~~~~~~  0.7&     2082&     2626&     3171\\
  0.50 --  0.75&  0.9&  0.7~~~~~~~~~  0.9&     1118&     1272&     1425\\
  0.75 --  1.00&  1.0&  0.7~~~~~~~~~  0.9&      730&      831&      931\\
  1.00 --  1.50&  1.0&  0.8~~~~~~~~~  1.0&      416&      450&      483\\
  1.50 --  2.00&  1.0&  0.8~~~~~~~~~  1.0&      215&      233&      250\\
  2.00 --  2.50&  1.0&  0.8~~~~~~~~~  0.8&      274&      274&      274\\
  2.50 --  3.00&  1.0&  0.8~~~~~~~~~  0.8&      227&      227&      227\\
  3.00 --  6.50&  1.0&  0.5~~~~~~~~~  0.5&       28&       28&       28\\
\enddata
\end{deluxetable}